\documentclass[aps,pra,twocolumn,nofootinbib,tightenlines,eqsecnum,showpacs,floatfix]{revtex4}
\usepackage{pslatex}
\usepackage{amsmath}
\input epsf
\numberwithin{equation}{section}
\renewcommand{\theequation}{\arabic{section}.\arabic{equation}}

\begin{document}
\newcommand{\ltwid}{\mathrel{\raise.3ex\hbox{$<$\kern-.75em\lower1ex\hbox{$\sim$}}}}
\newcommand{\gtwid}{\mathrel{\raise.3ex\hbox{$>$\kern-.75em\lower1ex\hbox{$\sim$}}}}

\title{Linear Positivity and Virtual Probability}

\author{James B.~Hartle}

\email{hartle@physics.ucsb.edu}

\affiliation{Department of Physics\\
University of California, Santa Barbara, CA 93106-9530}

\date{\today}

\begin{abstract}

We investigate the quantum theory of closed systems based on the linear
positivity decoherence condition of Goldstein and Page. The objective of
any quantum theory of a closed system, most generally the universe, is the
prediction of probabilities for the individual members of sets of
alternative coarse-grained histories of the system. Quantum interference 
between members of a set of alternative histories is
an obstacle to assigning probabilities that are consistent
with the rules of probability theory. 
A quantum theory of closed systems therefore requires two
elements; 1) a condition specifying which sets of histories may
be assigned probabilities, and 2) a rule for those probabilities. 

The linear positivity condition of Goldstein and Page is
the weakest of the general conditions proposed so far. Its general
properties relating to exact probability sum rules, time-neutrality, and
conservation laws are explored. Its inconsistency with the usual
notion of independent subsystems in quantum mechanics is reviewed.
Its relation to the stronger condition of medium decoherence necessary for 
classicality is discussed.  The linear positivity of histories in a number 
of simple
model systems is investigated with the aim of exhibiting linearly positive
sets of histories that are not decoherent. The utility
of extending the notion of probability to include values outside the range
of 0 to 1 is described. Alternatives with such virtual probabilities
cannot be measured or recorded, but can be used in the intermediate steps
of calculations of  real probabilities.  Extended probabilities give a simple
and general way of formulating quantum theory.

The various decoherence conditions are compared in terms of their utility for
characterizing classicality and the role they might play in further
generalizations of quantum mechanics.

\end{abstract}

\maketitle

\section{Introduction}

The familiar ``Copenhagen'' quantum mechanics of measured subsystems must
be generalized to apply to closed systems such as the universe as a whole.
That is because there is nothing outside a closed system to measure it.
Rather measurement situations occur 
within a closed system containing both the subsystem observed, and the
observers, apparatus, etc.~observing it. Consistent (or
decoherent) histories quantum theory provides such a 
generalization.\footnote{See {\it e.g.} \cite{Gri02,Omn94,Gel94} for
extended descriptions from various points of view.} 
This paper is concerned with the nature of the
decoherence
conditions which are central to these formulations of quantum theory.

The most general predictions of a quantum theory of a closed system are the
probabilities of coarse-grained alternative histories from the system's
initial quantum state and Hamiltonian.  The probabilities for alternative
orbits of the Earth around the Sun are examples.  An orbit might be
described by a series of center-of-mass position intervals 
at a sequence of times.  That description is coarse-grained because it is
specified only by center of mass position and not all possible variables,
because those positions are specified only at some times and not at all
times, and because they are specified only within intervals and not to
arbitrary accuracy.  Such coarse-graining is generally necessary to have
probabilities at all.  

Copenhagen quantum mechanics predicts the
probabilities of histories of measurement outcomes, and the quantum mechanics
of closed systems predicts these also.  However, these generalizations of
Copenhagen quantum mechanics also predict
probabilities for histories of subsystems even when these are not receiving
attention from observers. The history of the Earth's motion in the first
few billion years of its existence is an example.

Not every set of histories that can be described can be assigned
probabilities that are consistent with the rules of probability theory, in
particular, with the rule that the probability of two exclusive
alternatives 
is the sum of the probabilities of each.  Quantum interference is the 
obstacle.  In usual quantum mechanics, for instance, probabilities are 
the squares of amplitudes and the square of a sum is not generally the 
sum of the squares.  A quantum
theory of closed systems can therefore be based on two elements: 1) An
expression for computing probabilities and 2) a consistency  condition 
that specifies which sets of
alternative histories can be consistently assigned probabilities.

A variety of conditions enforcing consistency have been proposed and
studied (See, \textit{e.g.} \cite{GH90b} for a list). This paper is
concerned with the weakest of these --- linear positivity --- introduced in
a seminal paper by Goldstein and Page \cite{GP95}. Linear positivity will
be described in Sections II and III in detail, but to introduce the idea
let us first recall the medium decoherence condition.

Individual histories in an exhaustive set of exclusive alternative
histories are represented by operators $C_\alpha$ called
{\it class operators} where $\alpha$ labels the histories in the set. 
For instance, the history in which the orbit
of the Earth lies in a sequence of ranges of position at a series of times would
be represented by a $C_\alpha$ which is the time-ordered product of
projections onto these ranges at the different times.  The class operators
for an exhaustive set of histories satisfy
\begin{equation}
\sum\nolimits_\alpha C_\alpha = I\ , \qquad ({\rm exhaustive}).
\label{oneone}
\end{equation}
If $|\Psi\rangle$ is the state of the closed system, the class operators
can be used to construct {\it branch state vectors} $|\Psi_\alpha\rangle$
for each history according to
\begin{equation}
|\Psi_\alpha\rangle \equiv C_\alpha |\Psi\rangle.
\label{onetwo}
\end{equation}
The probabilities of the individual histories are given by
\begin{equation}
p (\alpha) = \|\, |\Psi_\alpha\rangle \|^2 = \| C_\alpha |\Psi\rangle\|^2.
\label{onethree}
\end{equation}
Medium decoherence is the condition that the branches be
orthogonal
\begin{equation}
\begin{split}
\langle\Psi_\alpha|\Psi_{\alpha^\prime}\rangle =&
\left\langle\Psi\left|C^\dagger_\alpha C_{\alpha^\prime}\right|
\Psi\right\rangle =0\quad \textrm{all} \quad \alpha\not= \alpha^\prime\ ,
\\
&({\rm Medium\ Decoherence})
\end{split}
\label{onefour}
\end{equation}
Medium decoherence is the precise representation of the absence of
interference between histories.  It is sufficient for the consistency of
the probabilities \eqref{onethree}. However, it is not a necessary
condition; weaker conditions are possible.

As a consequence of \eqref{oneone}, \eqref{onethree}, and \eqref{onefour},
the probabilities of a medium decoherent set of histories can be
re-expressed as
\begin{equation}
p (\alpha) \equiv \langle\Psi_\alpha|\Psi_\alpha\rangle =
\sum\nolimits_{\alpha^\prime} \langle\Psi_{\alpha^\prime}  |\Psi_\alpha
\rangle =
\langle\Psi|\Psi_\alpha\rangle = \langle\Psi|C_\alpha|\Psi\rangle.
\label{onefive}
\end{equation}
The numbers $\langle\Psi|C_\alpha|\Psi\rangle$ are real and positive as a
consequence of medium decoherence \eqref{onefour}. Goldstein and Page
\cite{GP95} proposed the weaker decoherence condition
\begin{equation}
{\rm Re}\ \langle\Psi|C_\alpha|\Psi\rangle \geq 0\ , \quad {\rm all}
\ \ \alpha, \qquad 
({\rm Linear\ Positivity}),
\label{onesix}
\end{equation}
replacing \eqref{onefour}, and a new assignment of probability
\begin{equation}
p (\alpha) \equiv {\rm Re}\ \langle\Psi|C_\alpha|\Psi\rangle,
\label{oneseven}
\end{equation}
replacing \eqref{onethree}.  This is {\it linear positivity}. ``Linear''
refers to the linearity of the expression for probabilities \eqref{oneseven}
in the $C_\alpha$.  ``Positivity'' characterizes the condition
\eqref{onesix}.  As we will review in Section III, the linear positivity
condition \eqref{onesix} is sufficient for the probabilities defined by
\eqref{oneseven} to be consistent.

The reality and positivity of the $\langle\Psi|C_\alpha|\Psi\rangle$ that
follow from \eqref{onefive} show that medium decoherence implies linear
positivity but not the other way around. Linear positivity thus
extends the sets of histories that can be assigned probabilities
beyond those which are medium decoherent. That extension is not
necessary to describe the outcomes of measurements. Sets of
histories describing measurements are medium decoherent, at least to
an excellent approximation, as we discuss in Section III.A.   

Linear positivity is the weakest
of the general conditions for consistent probabilities of histories known 
to the author at the time of writing. However mere consistency with
probability sum rules may not be the only requirement reasonably
imposed on quantum mechanical probabilities. For example, Diosi
\cite{Dio94} pointed out that linear positivity is inconsistent with
the usual notion of statistically independent subsystems in quantum
mechanics. The Hilbert space of a closed system of $N$,
non-interacting subsystems is the tensor product of the Hilbert
spaces for the individual subsystems. States representing
statistically independent subsystems have the product form 
\begin{equation}
|\boldsymbol{\Psi}\rangle = |\Psi_1\rangle \otimes \cdots \otimes |\Psi_N\rangle 
\label{tenprod}
\end{equation}
where the $|\Psi_i\rangle$ are the states of the individual
subsystems. The probability for the ensemble that the independent
subsystems have a  sequence of properties should be the product
of the probabilities of the properties in the individual subsystems.
While medium decoherence guarantees this property, linear
positivity does not, as Diosi showed and as we review in Section
III.D. For the author this is reason enough to reject linear
positivity as a basis for the quantum mechanics of closed systems
that contain isolated subsystems. 

However, usual quantum mechanics does not need to be generalized
for quantum cosmology only because the universe is a closed system. A
generalization is needed to accommodate quantum gravity. That is 
because even the quantum framework alluded to above, and described
more carefully in the next section, relies on a fixed background
spacetime geometry for example to define the notion of time employed.
But in a quantum theory of gravity spacetime geometry is a quantum
dynamical variable, generally fluctuating and without definite value.
The notion of an exactly isolated subsystem is problematical in 
a closed universe because the gravitational interaction cannot be
screened. It is therefore just possible that linear positivity 
could be useful could be useful for a generalization of quantum theory
incorporating quantum gravity. It is in this spirit that we 
explore linear positivity in this paper, although no gravitational 
issues will be addressed explicitly. 

This paper explores linear positivity in a number of different directions.
General properties such as exact probability sum rules, time neutrality,
conservation laws, the relation to medium decoherence, and the 
inconsistency with the usual notion of statistically independent 
subsystems are discussed in Section III. 
Section IV exhibits examples of linear
positivity in a variety of simple, non-relativistic quantum mechanical
systems --- the two-slit experiment, the three-box model, a spin-1/2
system, a free particle, and alternatives extended over time.  The aim is
to begin a kind of phenomenology of linear positivity.  In general, a much
wider range of histories of this simple kind are found to be linearly
positive than are decoherent. Section V describes the utility
of the numbers defined by \eqref{oneseven} even when they are not positive
or less than unity.  We call these virtual probabilities.  Alternatives
with virtual probabilities can neither be measured nor recorded but can be
used as intermediate steps in calculations of real probabilities.  We
discuss the advantages and disadvantages of formulating quantum
theory in terms of both in Section VI.

\section{Quantum Mechanics of Closed Systems}

This section very briefly reviews the essentials of the quantum mechanics
of closed systems \cite{Har93a} especially to fix the notation to be used in the
rest of the paper.

We consider a closed system, most generally the universe as a whole, in the
approximation that gross quantum fluctuations in the geometry of spacetime
(\textit{i.e.} quantum gravity) can be neglected.  The closed system can
then be thought of as a very large box of particles and fields containing
both observers and observed (if any). 
Fixed background spacetime means that a notion of time is well defined
and the usual apparatus of Hilbert space, states, and operators can be
employed in a quantum description of the box.  The quantum state of this
closed system $|\Psi\rangle$ (represented as a wave function of the 
universe) and the Hamiltonian $H$ describing its quantum evolution
are assumed to be given and fixed.

As mentioned in the Introduction, the most general objective of quantum
theory for a closed system is the prediction of probabilities for the
individual members of sets of coarse-grained alternative histories.  The
simplest sets of alternative histories are specified by alternatives at one
moment of time.  These can always be reduced to a set of yes/no
alternatives represented by an exhaustive set of orthogonal projection
operators $\{P_\alpha (t)\}$ $\alpha = 1, 2, \cdots$ in the Heisenberg
picture.  These projections satisfy
\begin{equation}
\sum\nolimits_\alpha P_\alpha (t) = I, \quad {\rm and} \quad P_\alpha
(t) P_\beta (t) = \delta_{\alpha\beta} P_\alpha(t)
\label{twoone}
\end{equation}
showing that they represent an exhaustive set of exclusive alternatives.
The operators $P_\alpha(t)$ change with time according to the Heisenberg
equation of motion defined by $H$. The state $|\Psi\rangle$ is unchanging
in the Heisenberg picture.

A set of alternative histories may be specified by giving sets of
alternatives $\{P^1_{\alpha_1} (t_1)\}$, $\{P^2_{\alpha_2} (t_2)\}, \cdots,
\{P^n_{\alpha_n} (t_n)\}$ at a series of times $t_1 < t_2 , \cdots, < t_n$.
The sets at distinct times can differ and are distinguished by the
superscript on the $P$'s.  The subscript distinguishes the alternatives
within the set. An individual history in this set is represented
by a particular sequence of alternatives $(\alpha_1, \cdots, \alpha_n)\equiv
\alpha$ and is represented by the corresponding chain of projections
\begin{equation}
C_\alpha = P^n_{\alpha_n} (t_n) \cdots P^1_{\alpha_1} (t_1).
\label{twotwo}
\end{equation}
Such a set of histories is coarse grained because alternatives are
specified at some times and not at every time and because the specified
alternatives are projections on subspaces with dimension larger than
one and not projections on a complete set of operators. 

The operations of coarse and fine graining relate different sets of
histories.  A set of  histories $\{C_\alpha\}$ may be coarse grained by partitioning
it into an exhaustive set of exclusive classes $\bar c_{\bar \alpha},
\bar\alpha=1, 2, \cdots$. Each class consists of some number of histories
in the finer-grained set and every finer-grained history lies in some
class.  Fine graining is the inverse operation of dividing the histories
up into mutually exclusive smaller classes.  The class operators 
$\{\bar C_{\bar\alpha}\}$ for
the histories in a coarse graining of a set whose class operators are 
$\{C_\alpha\}$ are related by summation  
\begin{equation}
\bar C_{\bar\alpha} = \sum\limits_{\alpha\in\bar\alpha} C_\alpha
\label{twothree}
\end{equation}
where the sum defining the coarse-grained history $\bar C_{\bar\alpha}$ 
is over all the
finer-grained histories contained within it.

Probabilities for a set of histories are a set of numbers $p(\alpha)$ lying
between 0 and 1
\begin{subequations}
\label{twofour}
\begin{equation}
0\leq p(\alpha) \leq 1
\label{twofoura}
\end{equation}
that satisfy the basic probability sum rules relating finer- to 
coarser-grained
descriptions: The probability of a coarse-grained set is the sum of the
probabilities of its finer-grained members.  The probabilities of the 
coarser-grained
descriptions are then {\it consistent} with those of the finer-grained
ones. Applied to histories for a coarse graining of the form \eqref{twothree}
this is the requirement  
\begin{equation}
\bar p (\bar\alpha) = \sum\limits_{\alpha\in\bar\alpha} p(\alpha)
\label{twofourb}
\end{equation}
\end{subequations}
where the $\bar p (\bar\alpha)$ are the probabilities of the $\bar
C_{\bar\alpha}$, and $p(\alpha)$ are the probabilities of the $C_\alpha$.
The two conditions \eqref{twofour} are the mathematical requirements for
probabilities.

The Introduction described how quantum interference is an obstacle to
consistency in the sense of \eqref{twofourb} and the need for a decoherence
condition to restrict the sets to ones where it is satisfied.  The next
Section describes the Goldstein and Page \cite{GP95} linear positivity
condition and some of its general
properties.

\section{Linear Positivity}
\label{III}

The linear positivity condition concerns the numbers
\begin{equation}
p(\alpha) \equiv {\rm Re}\, \left\langle\Psi|C_\alpha|\Psi\right\rangle = \frac{1}{2}
\left\langle\Psi\left|\left(C_\alpha + C^\dagger_\alpha\right)\right|\Psi
\right\rangle
\label{threeone}
\end{equation}
which can be calculated for any set of histories $\{C_\alpha\}$. It's
convenient to call these {\it candidate probabilities} and use the notation
$p(\alpha)$ for candidate probabilities whether or not their range is
between 0 and 1. Because of
\eqref{twothree}, they automatically and exactly satisfy the necessary sum
rules \eqref{twofourb} for the probabilities of a coarse graining of the
set.  However, candidate probabilities do not necessarily lie between 0 and
1 as \eqref{twofoura} requires. Indeed, even for histories with just two
times, with class operators of the form $C_{\alpha_2\alpha_1} =
P^2_{\alpha_2} (t_2)\, P^1_{\alpha_1} (t_1)$, there is some state in which
one of the histories has a negative $p(\alpha)$.  That is because the
Hermitian product of two non-commuting projections always has at least one
negative eigenvalue (Appendix A).

Homogeneous histories whose class operators are chains of projections as in
\eqref{twotwo} have candidate probabilities that are less than unity.  That
is because the action of a non-trivial projection on a state always reduces its
norm.
\begin{equation}
{\rm Re}\ \left\langle\Psi\left|P^n_{\alpha_n} (t_n) \cdots P^1_{\alpha_1} 
(t_1) \right|\Psi\right\rangle \leq 1.
\label{threetwo}
\end{equation}
There is no guarantee that a coarse graining of a set of
homogeneous histories that does not itself consist of homogeneous histories
will have $p(\alpha) < 1$. Indeed, for any set of homogeneous histories
with at least one negative $p(\alpha)$, there is always a coarse graining in 
which at least one history has $\bar p (\bar \alpha) \geq 1$. However, if the $p(\alpha)$
are all positive or zero then they are also less than $1$ because they must sum
to $1$ as a consequence of \eqref{oneone}. Genuine probabilities are therefore
guaranteed by the linear positivity condition \eqref{onesix}.


Section IV will illustrate the linear positivity condition for simple
systems. But first we consider some of its general properties and in
particular its relation to medium decoherence.

\subsection{Connection with Medium Decoherence}

A set of histories medium decoheres when
\begin{equation}
\left\langle\Psi\left|C^\dagger_\alpha C_\beta\right|\Psi\right\rangle
= 0, \quad \alpha \not= \beta \quad ({\rm Medium\ Decoherence}).
\label{threefour}
\end{equation}
The probabilities of the individual histories in a medium decoherent set is
\begin{equation}
p^{\rm MD} (\alpha) = \left\langle\Psi\left|C^\dagger_\alpha
C_\alpha\right|\Psi\right\rangle = \| C_\alpha |\Psi\rangle \|^2.
\label{threefive}
\end{equation}
As here, we use superscripts MD and LP where necessary to distinguish medium
decoherent and linear positivity probabilities.

The identity $\sum_\beta C_\beta=I$ [{\it cf.} \eqref{oneone}] can be used to
derive the following relation between the $p^{\rm LP}(\alpha)$ given by
\eqref{threeone} and the $p^{\rm MD}(\alpha)$ given by \eqref{threefive}
\begin{equation}
p^{\rm LP} (\alpha) = p^{\rm MD}(\alpha) + \sum\limits_{\beta\not=\alpha}
{\rm Re}\ \langle\Psi|C^\dagger_\beta
C_\alpha|\Psi\rangle.
\label{threesix}
\end{equation}
Evidently, the two notions of probability coincide if the set is exactly 
medium decoherent and the last term vanishes [{\it cf.} \eqref{threefour}].
Exact medium decoherence implies linear positivity.

However, many realistic coarse grainings are unlikely to medium decohere 
exactly, but only to an excellent approximation.  For
example, a branch state vector defining coarse-grained history of the
motion of the Earth to an accuracy of 1cm every second is not exactly
orthogonal to a branch state vector of a distinct history. Rather it is
approximately orthogonal to an accuracy far beyond that
to which the resulting probabilities can be checked or the physical
situation modeled.  Approximate medium decoherence does not necessarily
imply linear positivity because the second term in \eqref{threesix} may
have any sign.  If the $p^{\rm MD}(\alpha)$ are very small, $p^{\rm
LP}(\alpha)$ could be negative. (But then, approximate medium decoherence would 
imply approximate linear positivity.)

Medium decoherence can be alternatively characterized by the existence of
{\it records} of the histories.  There are records of a set of
histories if there is an
exhaustive set of orthogonal commuting {\it projections} 
$\{R_\beta\}$ which are
correlated with the histories $\{C_\alpha\}$ in the following strong sense
\begin{equation}
R_\beta C_\alpha|\Psi\rangle = \delta_{\beta\alpha} C_\alpha |\Psi\rangle.
\label{threeseven}
\end{equation}
Evidently \eqref{threeseven} implies the medium decoherence condition
\eqref{threefour}, because of the orthogonality of the $R$'s. Conversely,
if the branches $|\Psi_\alpha\rangle=C_\alpha |\Psi\rangle$ are orthogonal, 
there
are generally many different sets of projections $\{R_\alpha\}$ for which
\eqref{threeseven} is satisfied --- most simply $R_\alpha = |\Psi_\alpha\rangle
\langle\Psi_\alpha |$.
The $\{R_\alpha\}$ are records only in the abstract sense specified of
\eqref{threeseven} and need not correspond to anything like everyday
records such as history books.  The creation and persistence of records is
an important, indeed the defining, feature of many realistic mechanisms of
decoherence.

\subsection{Exact Probabilities}

As stressed by Goldstein and Page, and as mentioned above, the
probabilities of a linearly positive set of histories satisfy the defining
probability sum rules \eqref{twofourb} {\it exactly}.  Probabilities need
to be defined by physical theory only up to the accuracy they are used. For
example, we consider a prediction of the frequency of outcomes of  repeated 
experiments to be 
securely tested, not if the probability of a significant fluctuation 
in the frequency of expected outcomes is exactly zero, but rather if it
is sufficiently small.  Two theories that
predict probabilities whose difference is well below the standard with
which they are used are equivalent in predictive power.  Therein lies the
possibility of utilizing the probabilities of realistic approximately
medium decoherent sets of histories that obey the sum rules
\eqref{twofourb} only to an approximation secure beyond all test. However,
the conceptual situation is considerably simplified if the predicted
probabilities {\it exactly} obey the sum rules as they do in the linearly
positive case. We will return to the relation between these two cases in
Section VI.

\subsection{Exact Conservation Laws}

Exact probability sum rules ensure that quantities that commute with the
Hamiltonian are conserved with probability 1. The argument is essentially
the same as that in \cite{HLM95} but we provide a sketch of it here.  

Consider a quantity $A$ like total electric charge or total energy that
commutes with the Hamiltonian $H$. Let $P^{\rm A}_j (t)$ denote a set of
projections onto disjoint ranges $\Delta_j, j=1, 2, 3, \dots$ of $A$ at time $t$.
Orthogonality of these projections implies
\begin{equation}
P^{\rm A}_{j^\prime} (t^\prime) P^{\rm A}_j (t)=0, \quad j\not=j^\prime
\label{threeeight}
\end{equation}
because $A$ commutes with $H$. Now consider a set of histories of the
form
\begin{equation}
C_\alpha = P^{\rm A}_{j^\prime} (t^\prime) C^b_\beta P^{\rm A}_j (t)
\label{threenine}
\end{equation}
where $C^b_\beta$ is a  chain of projections at times in between $t$ and
$t^\prime$ onto ranges of quantities not necessarily commuting with $A$.  
Is the value of the conserved quantity $A$ at $t^\prime$ exactly
correlated with the value at $t$ despite the intervening projections on
quantities not necessarily commuting with $H$?  Linear positivity ensures
that it is.  Let $p(j^\prime, \beta, j)$ denote the candidate 
probabilities of the
histories \eqref{threenine} computed according to \eqref{threeone}. 
The sum rules \eqref{twofourb} ensure
\begin{equation}
\sum\nolimits_\beta p(j^\prime, \beta, j) = p(j^\prime, j) = 
\delta_{j^\prime j} p(j)
\label{threeten}
\end{equation}
from \eqref{threeeight}.  If the $p$'s are all positive as linear
positivity requires, then \eqref{threeten} implies
\begin{equation}
p(j^\prime, \beta, j)=0, \quad j^\prime\not= j
\label{threeeleven}
\end{equation} 
which is exact conservation of the quantity $A$.

\subsection{Statistical Probabilities}
\label{3D}

The probabilities under discussion are probabilities for particular
histories of a single closed system\footnote{See {\it e.g.} \cite{Har95c}
Section II.2 for a discussion of how these probabilities are used and
interpreted.}. When the closed system consists of an ensemble of identical
subsystems, the probabilities of the closed system can be used to discuss
the statistics of the ensemble when the number of its members is large.  For
example, with high probability, the frequency of a particular history in
the ensemble should equal its probability in an individual subsystem.  That
is a standard result in usual quantum theory \cite{probsum} in accord with the law
of large numbers. However, as 
observed by Diosi \cite{Dio94}, linear positivity is generally inconsistent with the
usual notion of statistical independence in quantum theory on which such
demonstrations are based. We offer a brief review of his result. 

In familiar quantum theory, an ensemble of $N$ identical, non-interacting
subsystem is represented by a state
\begin{equation}
|\boldsymbol{\Psi}\rangle = |\Psi\rangle \otimes \cdots \otimes
|\Psi\rangle
\label{threetwelve}
\end{equation}
on the tensor product of $N$ copies of the Hilbert space ${\cal H}$ of the
individual subsystems. A closed system with a state \eqref{threetwelve} can be
thought of as an approximate model for a realistic ensemble of identical
subsystems. Alternatively, even when $|\Psi\rangle$ refers to the universe,
\eqref{threetwelve} can be thought of as describing a fictitious ensemble of
universes whose frequencies can be expected to coincide with the
probabilities of the individual members in the limit of large $N$.

Let $\{C_\alpha\}$ denote the class operators acting on ${\cal H}$ for a
set of alternative, coarse-grained histories of a subsystem.  The class
operator for $N$ different histories $\alpha^1, \cdots, \alpha^N$ in the
ensemble is
\begin{equation}
{\bf C}_{\alpha^1 \cdots \alpha^N} = C_{\alpha^1} \otimes \cdots \otimes
C_{\alpha^N}.
\label{threethirteen}
\end{equation}
[We employ a superscript to distinguish histories $\alpha^1, \cdots,
\alpha^N$ of different subsystems each of which may consist of a sequence
of alternatives {\it e.g.} $\alpha^1=(\alpha^1_1, \cdots,
\alpha^1_n)$; $\alpha^2=(\alpha^2_1, \cdots, \alpha^2_n$);
etc.] Suppose that the set of histories $\{C_\alpha\}$ of a subsystem is
medium decoherent \eqref{threefour} so that in particular [{\it cf.}
\eqref{threefive}]
\begin{equation}
p^{\rm MD} (\alpha) = \|C_\alpha |\Psi\rangle \|^2.
\label{threefourteen}
\end{equation}
Then for the ensemble, evidently
\begin{equation}
\begin{split}
{\bf p}^{\rm MD} \left(\alpha^1, \cdots, \alpha^N\right) & =  
\left\| {\bf C}_{\alpha^1 \cdots \alpha^N} |\boldsymbol{\Psi}\rangle
\right\|^2, \\ &  = \left\|C_{\alpha^1} |\Psi\rangle \right\|^2 \cdots 
\left\|C_{\alpha^N} | \Psi\rangle \right\|^2,\\
& =  p^{\rm MD} (\alpha^1) \cdots p^{\rm MD} (\alpha^N).
\end{split}
\label{threefifteen}
\end{equation}
The subsystems are therefore statistically independent. This is the central
fact in deriving the result that for a very large ensemble the frequency of
an individual history $C_\alpha$ approaches its probability $p^{\rm MD}
(\alpha)$ in an individual subsystem.

However, linear positivity does not yield the same notion of statistical
independence. That is because [{\it cf.} \eqref{threeone}]
\begin{equation}
\begin{split}
{\bf p}^{\rm LP} (\alpha^1, \cdots, \alpha^N) & = {\rm Re}\left[
\left\langle\boldsymbol{\Psi}|{\bf C}_{\alpha^1 \cdots \alpha^N}
|\boldsymbol{\Psi}\right\rangle\right],\\
& =  {\rm Re} \left[\langle\Psi|C_{\alpha^1}|\Psi\rangle \cdots 
\langle\Psi|C_{\alpha^N}|\Psi\rangle\right].
\end{split}
\label{threesixteen}
\end{equation}
But the real part of a product is not generally the product of the real
parts unless the individual factors are all real. Thus
\begin{equation}
{\bf p}^{\rm LP} (\alpha^1, \cdots, \alpha^N) \not= p^{\rm LP} (\alpha^1)
\cdots p^{\rm LP} (\alpha^N).
\label{threeseventeen}
\end{equation}

The absence of general equality in \eqref{threeseventeen} does not mean
that linear positivity incorrectly predicts the frequencies of the histories
of the outcomes of measurements carried out on ensembles of idential subsystems of
the universe. Exactly measured or exactly recorded alternatives are exactly 
medium decoherent and then \eqref{threeseventeen} becomes an equality [{\it
cf.}\eqref{threefifteen}]. However, for many realistic sets of alternative 
histories, including those describing the quasiclassical realm of everyday
experience, medium decoherence and records are correlated with the
alternatives they record only to excellent approximations. In such cases
the \eqref{threeseventeen} will only be an equality to a related approximation.

There is a second reason that linear positivity does not incorrectly predict the frequencies of individual histories in an ensemble of identical subsystems.   Linear positivity fails for
ensembles with a sufficiently large number of identical subsystems even if it is satisfied
for the individual members unless the $\langle \Psi| C_\alpha|\Psi\rangle$ are real and \eqref{threeseventeen}
satisfied as consequence. To see this, consider a set of histories of the subsystem with just two
members:  $C_1$ and $C_2 = I - C_1$ (so that \eqref{oneone} is satisfied.) Write the
expected value of $C_1$ in terms of its magnitude and phase
\begin{equation}
\langle \Psi | C_1 | \Psi \rangle \equiv A  e^{i\phi}   
\label{threeeighteen1}
\end{equation}
with $A$ real and positive and $\phi$ real. Linear positivity requires $0<\phi<\pi$.

A given history of an ensemble of $N$ of these subsystems will have some number $n_C$
($0\le n_C \le N$) of histories $C_1$ and $N-n_C$ histories $C_2 = I -C_1$. 
The candidate probabilities $p(n_C)$ for histories of the ensemble with $n_C$ individual
histories $C_1$ are
\begin{equation}
p(n_C)= {\rm Re}\left[ (A e^{i\phi})^{n_C} (1- A
e^{i\phi})^{N-n_C}\right] \ .  
\label{insert1}
\end{equation}
For the histories of the ensemble to be linearly positive these candidate probabilities must
be positive for {\it all} $n_C$ between $0$ and $N$. However, unless $\phi$ is identically
zero, there is some large $N$ for which \eqref{insert1} will not be positive for some $n_C$.
Linear positivity thus fails for sufficiently large ensembles unless the $\langle \Psi| C_\alpha|\Psi\rangle$ are real .
The failure means probabilities cannot be assigned to individual histories of the ensemble 
much less to values of the frequency of any history of its individual subsystems\footnote{
The author thanks S. Goldstein for stressing this point.}.

A decoherence condition that would ensure equality in \eqref{threeseventeen} and
consistency with the usual notion of independent subsystems is 
\begin{subequations}
\label{threeeighteen}
\begin{align}
&{\rm Im}\ [\langle\Psi|C_\alpha|\Psi\rangle] = 0 
\label{threeeighteen a}, \quad {\rm all}\ \alpha,  \\
&{\rm Re}\ \left\langle\Psi\left|C_\alpha\right |\Psi\right\rangle 
\ge 0, \quad {\rm all} \ \alpha,  \label{threeeighteen b}\\
&({\rm Real\ Linear\ Positivity}) . \nonumber
\end{align}
\end{subequations}
This is linear positivity \eqref{onesix} augmented by the reality  requirement  \eqref{threeeighteen a}. 
At the risk of introducing more confusing terminology this might be called {\it real 
linear positivity}.  This is weaker than
medium decoherence because \eqref{onefive} and \eqref{onetwo} show that it
is equivalent to 
\begin{equation}
{\rm Im}\ \sum_{\beta} \left[\langle\Psi|C^\dagger_{\beta}
C_\alpha| \Psi\rangle\right] = 0.
\label{threenineteen}
\end{equation}
Only the vanishing of the imaginary part of this  sum is required for equality in
\eqref{threeseventeen},
but medium decoherence ensures that each term with $\alpha\ne\beta$ vanishes separately.  But, as
with medium decoherence, \eqref{threenineteen} is likely to be satisfied only to
an excellent approximation for realistic coarse grainings defining the
quasiclassical realm. 

The conflict between the usual notion of independent subsystems and linear
positive probabilities is a serious obstacle to their interpretation when
they are not associated with medium decoherent sets of histories. For
example, we cannot understand them as frequencies in an imaginary ensemble
identical copies of the closed system as described above. 
Indeed, it would not be unreasonable
to impose \eqref{threefifteen} as a {\it condition} for a  construction of
probabilities in quantum mechanics thereby ruling out linear positivity.
However, all this does not mean linear positive probabilities cannot be
useful as we discuss in Section VI.

\subsection{Time Symmetry and Asymmetry}

Quantum theory is usually formulated with an arrow of time\footnote{See,
{\it e.g.} \cite{GH93b} for a fuller discussion in the notation of this
paper.}.  Nowhere is
that seen more clearly than in the standard formula 
[{\it cf.}~\eqref{threefive}] for the probability $p^{\rm
MD}(\alpha_n,\cdots,\alpha_1)$ of a history of alternatives $\alpha_1,
\cdots, \alpha_n$ at a sequence of times $t_1<t_2<\cdots <t_n$.
\begin{equation}
p^{\rm MD} (\alpha_n, \cdots, \alpha_1) = \| P^n_{\alpha_n} (t_n) \cdots
P^1_{\alpha_1} (t_1) |\Psi\rangle \|^2.
\label{threetwenty}
\end{equation}
The state occurs at one end of the histories; there is
nothing at the other end. That is  the quantum mechanical arrow of time. 
It is a convention that $|\Psi\rangle$ enters as an {\it initial}
condition earlier than all the alternatives.  Field theory is CPT
invariant and  utilizing a CPT transformation the time order of the operators
could be inverted so that $|\Psi\rangle$ is a final condition.
However, no CPT transformation can alter the asymmetry between initial and
final conditions in \eqref{threetwenty}.

As discussed by Goldstein and Page, the linear positivity formula for
probabilities [{\it cf.} \eqref{threeone}] is time-neutral, {\it viz.}
\begin{equation}
\begin{split}
p^{\rm LP} (\alpha_n,\cdots, \alpha_1)& = {\rm Re} \left\langle\Psi\left|
P^n_{\alpha_n} (t_n) \cdots P^1_{\alpha_1} (t_1)\right|\Psi\right\rangle
\\
&= {\rm Re}\left\langle\Psi\left|P^1_{\alpha_1} (t_1) \cdots P^n_{\alpha_n}
(t_n) \right|\Psi\right\rangle.
\end{split}
\label{threetwentyone}
\end{equation}
The ends of histories are treated symmetrically.\footnote{There are
time-neutral formulations of medium decoherence involving both initial and
final conditions, {\it e.g.} \cite{GH93b}.}.

If the more general notion of linear positivity is fundamental in quantum 
theory, what is the
origin of the time asymmetry displayed by the medium decoherence expression
 \eqref{threetwenty}? The answer
is that the approximate equality of $p^{\rm LP} (\alpha_n, \cdots,
\alpha_1)$ and $p^{\rm MD} (\alpha_n, \cdots, \alpha_1)$ can be expected to
hold only for one ordering of the projections in \eqref{threetwenty}.
Put differently, the second term in
\eqref{threesix} is not time-neutral. It will generally be small only for
one-time ordering of the projections in $C_\alpha$ and not both.
Only in trivial cases, {\it e.g.} projections on
conserved quantities will the probabilities of an ordering and the inverse
ordering agree.

Records of history connect the quantum mechanical arrow of time in
\eqref{threetwenty} with the second law of thermodynamics for sets of
histories constituting the quasiclassical realm of everyday experience.
The histories of the quasiclassical realm \cite{GH90a,Har94b,GH93a,GHup} 
consist of projections on
ranges of the ``hydrodynamic'' variables of classical physics. These
generally are integrals over suitably small volumes of densities
of approximately conserved quantities such as energy, momentum, numbers of
different species, etc. 

The entropy of a set of histories consisting of
alternatives $\{P_\alpha (t)\}$ at a single moment of time is
\begin{equation}
\begin{split}
S(\{P_\alpha (t)\}) = &- \sum\nolimits_\alpha p(\alpha, t) 
\log p(\alpha, t) \\ & +
\sum\nolimits_\alpha p(\alpha, t) \log {\rm Tr}\, [P_\alpha (t)]
\end{split}
\label{threetwentytwo}
\end{equation}
where $p(\alpha, t) = \|P_\alpha (t)|\Psi\rangle\|^2$.  This coincides with
the usual entropy of physical chemistry and physics when the
$P_\alpha (t)$ are projections on the hydrodynamic quasiclassical variables 
described above \cite{GHup}. It is low initially for the state of our 
universe and therefore has a tendency to increase with $t$ afterward. 
That is the second law of thermodynamics.

Records of history in the quasiclassical realm are often created in
irreversible processes where the entropy defined above
increases\footnote{Increase in entropy is not {\it necessary} to create
a record as in reversible computation \cite{Lan61}.}.  An
impact crater on the Moon, an ancient fission track in mica, a darkened
photographic grain, and the printing of ink on this paper are all examples.
Consistent with the second law, records of events in history are more
likely to be at the end of history furthest from an initial condition of
low entropy rather than at the beginning.

The existence of records is another way of characterizing medium
decoherence as discussed in Section III.A. The second law suggests that
\eqref{threeseven} holds in the form given and not with $C_\alpha R_\beta$
on the left-hand side. This leads to the medium decoherence of $C_\alpha$
but not the set $\{C^\dagger_\alpha\}$ with the projections in the inverse
order.  Both have the same linear positive probabilities, but only one
order is connected with medium decoherence.  Irreversible creation of records
is therefore one reason why the quantum mechanical arrow of time is
consistent with the thermodynamic arrow of time, at least for the
quasiclassical realm.

\section{Examples of Linear Positivity}

This section calculates the candidate probabilities $p(\alpha) = {\rm Re}
\langle\Psi|C_\alpha|\Psi\rangle$ for a number of simple examples of
physical systems, states $|\Psi\rangle$, and sets of histories
$\{C_\alpha\}$.  The intent is not to be exhaustive but to illustrate
linearly positive sets of histories that are not necessarily decoherent.
It's not necessary to study all of these examples to understand the general
discussion of quantum theory that follows.

\subsection{The Two-slit Experiment}

We begin with the classic two-slit experiment shown in Fig.~1.  Electrons
from a source $S$ travel toward a vertical screen with two horizontal slits
each a distance $d/2$ from the axis perpendicular to the
screen through $S$.  The electrons are later detected at a second screen, parallel to
the first, a distance $D$ away.  A coordinate $y$ measures the vertical
distance from the axis to the point of detection.

\begin{figure}[t]
\centerline{\epsfysize=2in \epsfbox{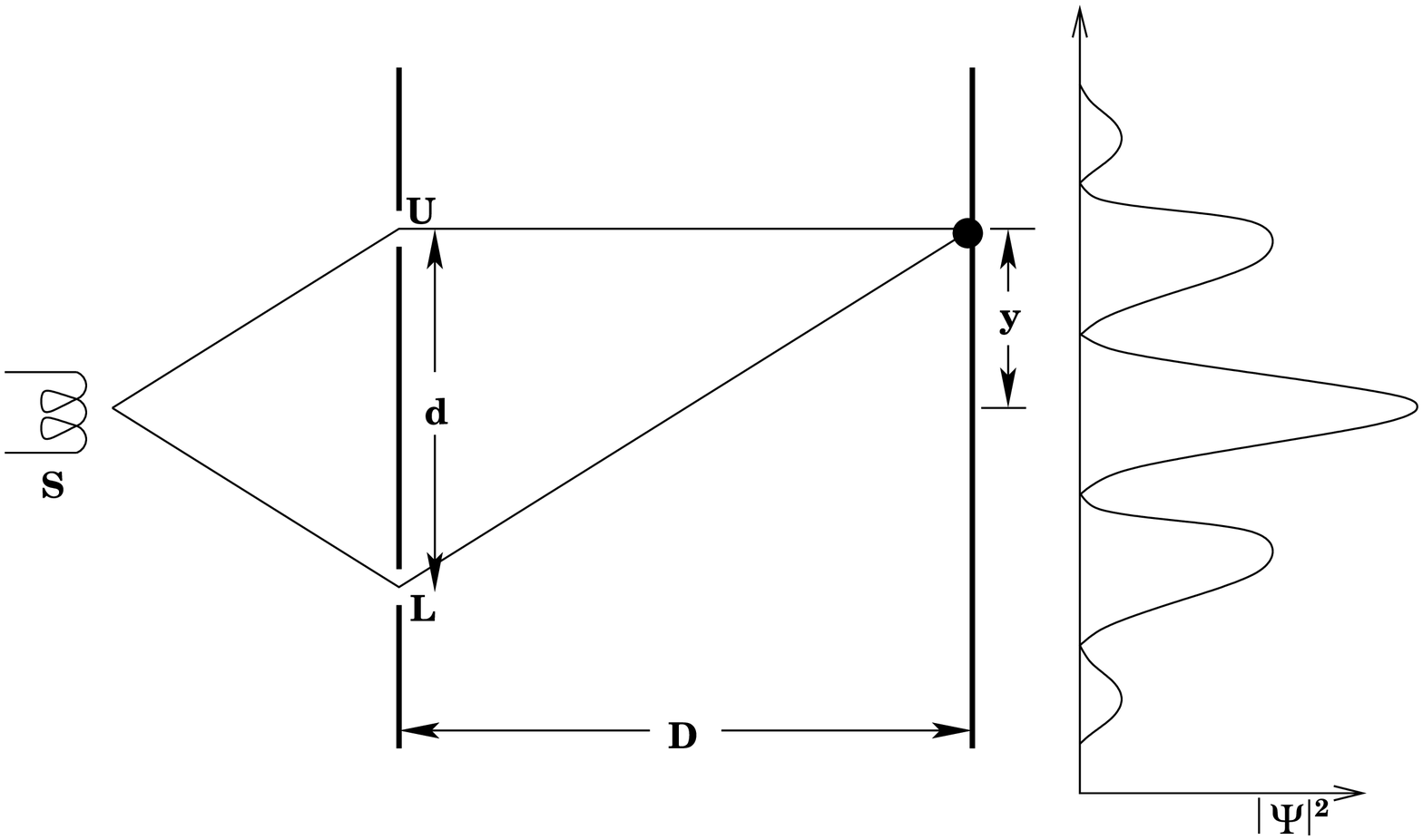}}
\caption{The Geometry of the two-slit experiment. An electron gun at left
emits an electron traveling toward a screen with two slits, its progress in
space recapitulating its evolution in time. The electron is detected on the
screen at right at a position $y$ with a probability density that exhibits
an interference pattern. A coarse-grained set of histories for the electron
can be defined by specifying which slit ($U$ or $L$) it went through and ranges
of detected positions $y$.}
\label{twoslit}
\end{figure}

We make the usual idealizations, in particular assuming that
the electrons are initially in narrow wave packets moving
in the horizontal direction $x$, so that their
progress in $x$ recapitulates their evolution in time.  We assume that the
source is far enough from the first screen that these wave packets can be
analyzed into plane waves propagating in the $x$ direction with a
distribution peaked about a wave number $k$. Then, to a good approximation, we
can calculate the amplitudes for detection by considering just the plane
wave with peak wave number $k$. That analysis follows. 

At the detecting screen the amplitude that the electron travels through the
upper slit $U$ to arrive at
a point $y$ on the detecting screen is
\begin{equation}
\Psi_U(y) \equiv ae^{ikS_U(y)}/S_U(y)
\label{fourone}
\end{equation}
where
\begin{equation}
S_U(y)\equiv \left[\left(d/2 -y\right)^2 + D^2\right]^{\frac{1}{2}}
\label{fourtwo}
\end{equation}
is the distance from the upper slit to the point on the detecting screen
labeled by $y$ and $a$ is a constant amplitude. Similarly, the amplitude 
to pass
through the lower slit $L$ and arrive at $y$ is
\begin{equation}
\Psi_L (y) \equiv a e^{ikS_L(y)}/S_L(y)
\label{fourthree}
\end{equation}
with
\begin{equation}
S_L(y) \equiv \left[\left(d/2  + y\right)^2 +
D^2\right]^{\frac{1}{2}}.
\label{fourfour}
\end{equation}

The candidate probability {\it densities} $\wp(y, U)$ 
and $\wp(y, L)$ to arrive at
$y$ in an interval $dy$ having passed through the upper or lower slit are
given by [\textit{cf.}~\eqref{threeone}]
\begin{subequations}
\begin{align}
&\wp (y, U) = {\rm Re}\ \left[\Psi^* (y) \Psi_U (y)\right], \\ &\wp (y,
L) = {\rm Re}\ \left[\Psi^* (y) \Psi_L (y)\right].
\label{fourfive}
\end{align}
\end{subequations}
where $\Psi(y) = \Psi_U (y) + \Psi_L (y)$.
The first of these works out to be 
\begin{equation}
\wp (y, U) = \frac{|a|^2}{S_U}\ \left\{\frac{1}{S_U} + \frac{1}{S_L}
\cos [k(S_L-S_U)]\right\}.
\label{foursix}
\end{equation}
The expression for $\wp (y, L)$ is the same with $L$ and $U$
interchanged. The probability density to arrive at $y$ irrespective of
which slit is passed through is
\begin{subequations}
\label{fourseven}
\begin{eqnarray}
\wp_{\rm tot} (y) & = & \wp (y, U) + \wp (y, L),\\
\label{foursevena}
& = & |a|^2 \left\{\frac{1}{S^2_U} + \frac{1}{S^2_L} + 2 \frac{\cos [k
(S_L-S_U)]}{S_U S_L}\right\}, \\
\label{foursevenb}
& = & |\Psi_U (y) + \Psi_L (y) |^2.
\label{foursevenc}
\end{eqnarray}
\end{subequations}
The last equality of course follows directly from \eqref{threeone}.  
Figure 2 shows
$\wp (y, U)$, $\wp (y, L)$, $\wp_{\rm tot} (y)$ for an apposite choice of
$D$ and $d$.

\begin{figure}[t!]
\centerline{\epsfysize=6in \epsfbox[86 94 399 707]{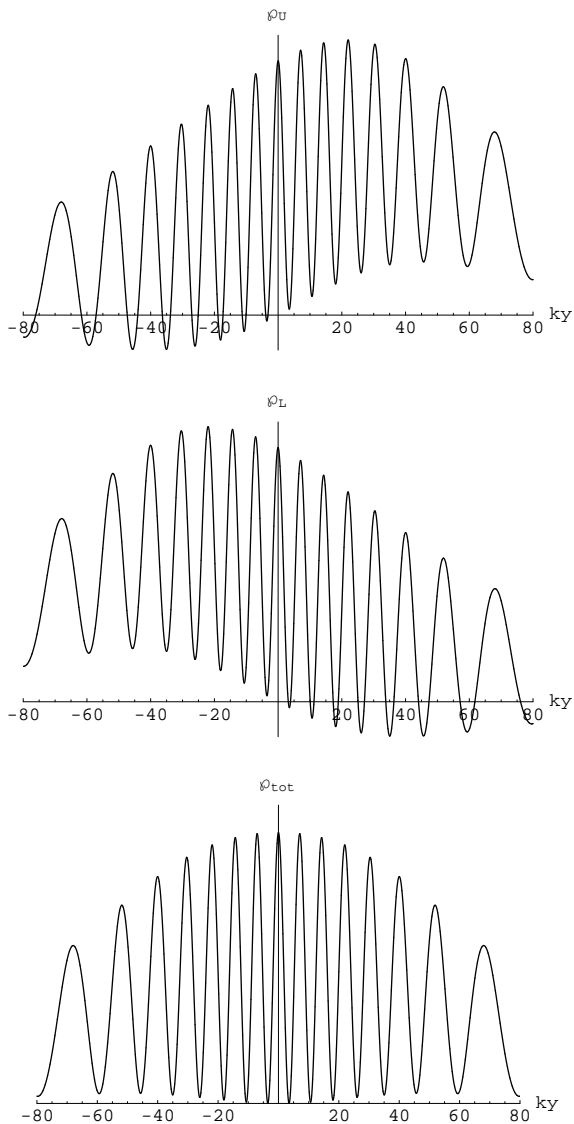}}
\caption{Candidate probability densities for the two-slit experiment. The
candidate probability density $\wp(y, U)$ to go through the upper slit
and arrive at $U$ is shown at top for $kd=kD=60$ where $k$ is the wave
number of the electron [{\it cf.} Fig.~1]. 
The corresponding candidate probability density $\wp (y,
L)$ for going through the lower slit is just below. The sum $\wp_{\rm tot}$
is the probability density to arrive at $y$ irrespective of which slit is
passed through is shown at bottom. The amplitude $a$ has been assigned
arbitrarily which is why the scale of the vertical axis not indicated.
When these probability densities are integrated over ranges of $y$, they
give candidate probabilities for the histories. The size of well-chosen
ranges that yield positive probabilities is smaller than the size that
would wash out the interference pattern.}
\end{figure}

A set of histories of the electrons can be defined by whether the electron
passed through the upper and lower slit and arrived in one or the other of
an exhaustive sets of exclusive ranges of $y$ of size $\{\Delta_\alpha\}$,
$\alpha$ an integer. The candidate probabilities for these histories are
the integrals of the densities $\wp (y, U)$ and $\wp (y,L)$ over
the ranges $\{\Delta_\alpha\}$, {\it e.g.}
\begin{equation}
p(\Delta_\alpha, U) = \int_{\Delta_\alpha} dy\, \wp (y, U).
\label{foureight}
\end{equation}

Consider for simplicity the candidate probabilities
illustrated in Fig.~2 with all the ranges
of equal size $\Delta$. Evidently if $\Delta$ is too small, the set of
histories will not be linearly positive because some of the candidate
probabilities \eqref{foureight} will be negative.  Equally evidently, if
$\Delta$ is sufficiently large the set will be linearly positive and the
$p$'s will be probabilities.  Most importantly however, the size $\Delta$ 
necessary
to achieve positivity is {\it smaller} than that needed to wash out the
interference pattern. Indeed in the limit $D\gg d$ the probability densities
are  
\begin{equation}
\wp (y, U) = \wp (y, L) \sim \frac{|a|^2}{D^2}
\ \left[1+\cos \left(\frac{kyd}{D}\right)\right].
\label{fournine}
\end{equation}
Since these are both positive, arbitrarily small $\Delta$ leads to positive
probabilities.

The set of histories will be medium decoherent to a good approximation
where $\Delta$ is sufficiently large that the interference term averages to
zero
\begin{equation}
\int_\Delta dy\, {\rm Re} \left[\Psi^*_L (y) \Psi_U (y)\right] \approx 0.
\label{fourten}
\end{equation}
That requires a $\Delta$ larger than the spacing between the interference
fringes.  As already mentioned, that is much larger than the $\Delta$
required for linear positivity showing concretely how
linear positivity is a weaker condition than weak decoherence. 

\subsection{Spin -- 1/2}

Consider a single free spin in a state $|\Psi\rangle$.  For simplicity,
assume that its Hamiltonian is zero.  Consider histories specified by
giving the value of the spin at one time  along direction $\vec n_1$
and at a later time along a direction $\vec n_2$.
Since the Hamiltonian is zero, it's not necessary to specify the particular
times, only the time ordering of the alternatives matters. 

Let $P^{\vec n}_s$ denote the projections on the two orientations 
of the spin along direction $\vec n$ where $s$ ranges over the possible
orientations $(+, -)$.  The branch state vectors are
\begin{equation}
|\Psi_{s_2s_1}\rangle = P^{\vec n_2}_{s_2} P^{\vec n_1}_{s_1} |\Psi\rangle
\label{foureleven}
\end{equation}
and the candidate probabilities are
\begin{equation}
p (s_2, s_1) = {\rm Re}\ \left\langle\Psi|\Psi_{s_2s_1}\right\rangle = 
{\rm Re}\ \left\langle\Psi\left| P^{\vec n_2}_{s_2} P^{\vec n_1}_{s_1}
 \right|\Psi\right\rangle.
\label{fourtwelve}
\end{equation}

To exhibit the candidate probabilities \eqref{fourtwelve}
 explicitly it's convenient to introduce a rectangular $(x,
y, z)$ coordinate system with $z$ oriented along $\vec n_2$ and $y$ chosen
so that $\vec n_1$ lies in the $y-z$ plane.  The angle between $\vec n_1$
and $\vec n_2$ we denote by $\delta$. In a general state $|\Psi\rangle$, the
spin points along some direction which we take to be specified by polar
angles $\theta$ and $\phi$ with respect to the $z$-axis.  Thus in a basis
in which $s_z$ is diagonal we can take
\begin{equation}
\Psi = \left(\begin{array}{c} e^{i\phi/2}\cos( \theta/2)\\
e^{-i\phi/2} \sin( \theta/2)\end{array}\right).
\label{fourthirteen}
\end{equation}
It is then straight forward to calculate the four candidate probabilities 
\eqref{fourtwelve}.
We find
\begin{subequations}
\label{fourfourteen}
\begin{align}
p(+, +) & =  \cos^2\left(\frac{\theta}{2}\right) \cos^2\left(\frac{\delta}{2}
\right) +\frac{1}{4} \cos \phi \sin
\theta  \sin \delta
\label{fourfourteena} \\
p(+, -) & =  \cos^2\left(\frac{\theta}{2}\right) \sin^2\left(\frac{\delta}{2}
\right) -\frac{1}{4} \cos \phi \sin\theta\sin\delta
\label{fourfourteenb} \\
p(-, +) & =  \sin^2\left(\frac{\theta}{2}\right) \sin^2\left(\frac{\delta}{2}
\right) +\frac{1}{4} \cos \phi \sin\theta\sin\delta
\label{fourfourteenc}\\
p(-, -) & =  \sin^2\left(\frac{\theta}{2}\right) \cos^2\left(\frac{\delta}{2}
\right)-\frac{1}{4} \cos \phi \sin\theta\sin\delta
\label{fourfourteend}
\end{align}
\end{subequations}
Note that these correctly sum to unity and that $p(+) = p(+, +) + p(+, -)$,
etc.

Figure 3 shows the domain of states parametrized by $(\theta, \phi)$ that
imply a linearly positive set of histories for various values of the angle
$\delta$. For each $\delta$ there is a significant range of states giving
linearly positive histories which shrinks as $\delta$ approaches 0 or $\pi$
even though those limits correspond to linear positivity and indeed exact
medium decoherence for all states.  Any state with the phase $\phi=\pi/2$
leads to a linearly positive set of histories for any $\delta$.

\begin{figure}[t!]
\centerline{\epsfysize=6in \epsfbox{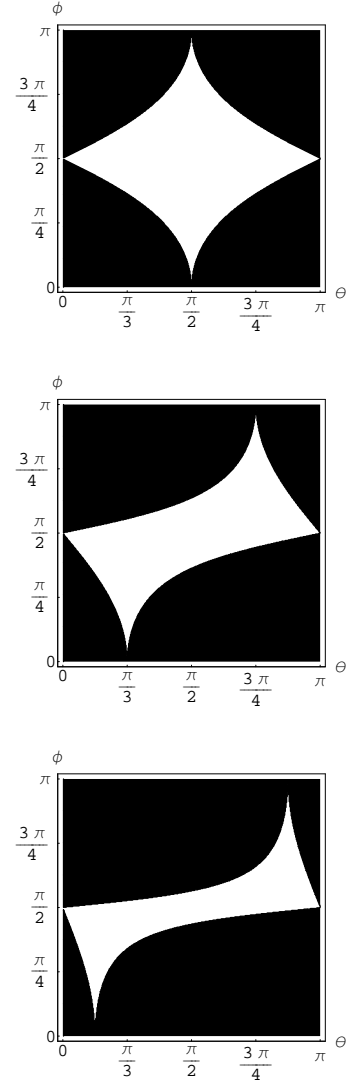}}
\caption{Candidate probabilities for two different spin directions of a
spin-1/2 particle. States of a spin-1/2 system can be labeled by two angles
$(\theta, \phi)$ as in \eqref{fourthirteen}. These plots are concerned with the
candidate probabilities for histories specified by values of the spin along
a direction $\vec n_1$ followed (or preceded) immediately by a value of the
spin along a second direction $\vec n_2$ making an angle $\delta$ with
respect to the first.  The plots show the states for which this set of
histories is linearly positive (white) and those for which some candidate
probabilities are negative (black) for given $\delta$. Only the partial range $0<\phi<\pi$ is shown
because the rest can be determined from the symmetries of \eqref{fourfourteen}. From top to bottom
the values of $\delta$ are $\pi/2$, $\pi/4$, and $\pi/8$. For example,
states with $\phi = \pi/2$ imply these sets are linear positive 
for any value of $\delta$ but none is medium decoherent.}
\end{figure}

Almost none of these linearly positive sets of histories are medium
decoherent with orthogonal branches.  The conditions for medium decoherence
that aren't automatically satisfied are the vanishing of the
following:
\begin{subequations}
\label{fourfifteen}
\begin{align}
&
\begin{split}
\left\langle\Psi_{++}|\Psi_{+-}\right\rangle  = & \frac{1}{4}\ \sin
\delta [+\cos \theta \sin \delta \\ 
& + \sin \theta (\cos \delta \cos \phi - i \sin \phi)], 
\label{fourfifteena} 
\end{split} \\ 
& \begin{split}
\left\langle\Psi_{--}|\Psi_{-+}\right\rangle = & \frac{1}{4}\ \sin
\delta [-\cos \theta \sin \delta  \\ 
&+ \sin \theta (\cos\delta \cos \phi + i \sin \phi)].
\label{fourfifteenb}
\end{split}
\end{align}
\end{subequations}
Both of these vanish only for $\theta = \pi/4$ or $3\pi/4$ and $\phi = 0$
or $\pi$. For example as mentioned above, any state with $\phi=\pi/2$
leads to linearly positive histories.  But none of these states implies
medium decoherence. Linear positivity is evidently a much weaker condition.  

\subsection{The Three-Box Example}

The three-box example introduced by Aharonov and Vaidman \cite{AV91} for
other purposes has proved useful for illustrating the nature of quantum
reality in consistent histories quantum mechanics \cite{Ken97,GrH97}.  
We use it here to
illustrate the scope of the linear positivity condition.

Consider a particle that can be in one of three boxes, $A$, $B$, $C$ in
corresponding orthogonal states $|A\rangle$, $|B\rangle$, and $|C\rangle$.
For simplicity, take the Hamiltonian to be zero, and suppose the system to
initially be in the state
\begin{equation}
|\Psi \rangle \equiv \frac{1}{\sqrt{3}}\ (|A\rangle + |B\rangle +
|C\rangle).
\label{foursixteen}  
\end{equation}
Consider further a state $|\Phi\rangle$ defined by 
\begin{equation}
|\Phi\rangle \equiv \frac{1}{\sqrt{3}}\ (|A\rangle + |B\rangle -
|C\rangle)
\label{fourseventeen}
\end{equation}
and denote the projection operators on $|\Phi\rangle$, $|A\rangle$,
$|B\rangle$, $|C\rangle$ by $P_\Phi$, $P_A$, $P_B$, $P_C$ respectively.
Denote with $\bar A$ the negation of $A$ (``not in box $A$'') represented
by the projection $P_{\bar A} = I-P_A$. The negations $\bar \Phi, \bar B,
\bar C$ and their projections $P_{\bar \Phi}, P_{\bar B}$, and $P_{\bar C}$
are similarly defined. We consider sets of histories, all of which specify
whether the particle is in state $|\Phi\rangle$ or not at a final time and in
various boxes at an intermediate time after the initial one.  The exact
values of these times is unimportant since $H=0$. Only the order matters
--- initial, intermediate, and final.  

An example is supplied by histories which specify whether the particle is
in box $A$ or not at the intermediate time.  There are four alternative
histories represented by the class operators
\begin{equation}
P_\Phi P_A\ ,\ P_\Phi P_{\bar A}\ , \ P_{\bar \Phi} P_A\ , \ P_{\bar\Phi}
P_{\bar A}.
\label{foureighteen}
\end{equation}
Their candidate probabilities are given by \eqref{threeone} {\it e.g.}
\begin{equation}
p(\Phi, A) = {\rm Re}\ \langle\Psi|P_\Phi P_A|\Psi\rangle.
\label{fournineteen}
\end{equation}
A little calculation  shows that
\begin{equation}
\begin{split}
 p(\Phi, A) = 1/9, \quad &p(\bar\Phi, A) = 2/9, \\ p(\Phi, \bar A) =0,\ \ \quad &p(\bar\Phi, \Bar A)= 2/3.
\end{split}
\label{fourtwenty}
\end{equation}
This is a linearly positive set of histories since all the numbers
\eqref{fourtwenty} are positive or zero.  Indeed, this set of histories is
medium  because the branch state vectors obtained by applying the class
operators \eqref{foureighteen} to $|\Psi\rangle$ [{\it
cf.}~\eqref{threefour}]
are all orthogonal.  The positivity of probabilities follows just from
that.

Next consider the finer-grained set of histories which specifies whether or
not the particle is in box $A$ {\it and} whether or not it is in box $B$ at the
intermediate time.  The eight class operators are
\begin{equation}
P_\Phi P_A P_B\ ,\ P_\Phi P_A P_{\bar B}\ ,\ P_\Phi P_{\bar A} P_B
\ ,\ \cdots, etc.
\label{fourtwentyone}
\end{equation}
This set of histories is {\it not} medium decoherent. The candidate probabilities
are
\begin{subequations}
\label{fourtwentytwo}
\begin{eqnarray}
p(\Phi, A, B) & = & p(\bar\Phi, A, B) = 0,
\label{fourtwentytwoa} \\
p(\Phi, A, \bar B) & = & p (\Phi, \bar A, B) = 1/9,
\label{fourtwentytwob} \\
p (\Phi, \bar A, \bar B) & = & - 1/9,
\label{fourtwentytwoc} \\
p(\bar\Phi, A, \bar B) & = & p(\bar\Phi, \bar A, B) = 2/9, 
\label{fourtwentytwod} \\
p (\bar\Phi, \bar A, \bar B) & = & 4/9.
\label{fourtwentytwoe}
\end{eqnarray}
\end{subequations}
The one negative number, $p (\Phi, \bar A, \bar B)= -1/9$, shows
that this non-decoherent set of histories is also not linearly positive.

A seeming contradiction occurs \cite{Ken97} when one calculates 
the conditional  
probabilities for the particle to be in box $A$ or $B$ {\it given} that
it is in state $\Phi$ at the later time.  These are
\begin{equation}
p(A|\Phi) = \frac{p(\Phi, A)}{p(\Phi)}\ ,\qquad p(B|\Phi) = \frac{p(\Phi,
B)}{p(\Phi)}.
\label{fourtwentythree}
\end{equation}
The symmetry between $A$ and $B$ in the definitions of $|\Psi\rangle$ and
$|\Phi\rangle$ in \eqref{foursixteen} and \eqref{fourseventeen} implies
$p(A|\Phi)$ and $p(B|\Phi)$ are equal. Calculation shows that they are both
unity.
But this would seem to be a contradiction because being in $A$ and being in
$B$ are exclusive alternatives: $P_A P_B\equiv 0$ implying $p(A, B) = p(A,
B|\Phi)=0$ from \eqref{threeone}.  From that one would like to infer from
$p(A|\Phi)=1$ that $p(B|\Phi)=0$. In fact, this inference cannot be
drawn \cite{GrH97} and there is no contradiction. Let us see how this
is established in the present context.

To calculate $p(B|\Phi)$ from $p(A|\Phi)$ and $p(A, B|\Phi)=0$, we need the
finer-grained set of histories \eqref{fourtwentytwo} referring both to box $A$
and $B$ and to $\Phi$. In this set it follows from \eqref{twofourb} that 
\begin{subequations}
\label{fourtwentyfour}
\begin{eqnarray}
p(A|\Phi) & = & p(A, B|\Phi) + p(A, \bar B|\Phi),\\
\label{fourtwentyfoura}
p(\bar A|\Phi) & = & p(\bar A, B|\Phi) + p(\bar A, \bar B|\Phi).
\label{fourtwentyfourb}
\end{eqnarray}
\end{subequations}
Similarly,
\begin{subequations}
\label{fourtwentyfive}
\begin{eqnarray}
p(B|\Phi) & = & p(A, B|\Phi) + p(\bar A, B|\Phi),
\label{fourtwentyfivea} \\
p(\bar B|\Phi) & = & p(A, \bar B|\Phi) + p(\bar A, \bar B|\Phi).
\label{fourtwentyfiveb}
\end{eqnarray}
\end{subequations}
If $p(A|\Phi)=1$ then $p(\bar A|\Phi)=0$. Were the $p$'s positive
probabilities, equating \eqref{fourtwentyfourb} to zero would imply that
both $p(\bar A, B|\Phi)$ and $p(\bar A, \bar B|\Phi)$ were zero. Inserting
the first of these in \eqref{fourtwentyfivea} along with $p(A, B|\Phi)=0$
from \eqref{fourtwentytwoa} gives $p(B|\Phi)=0$.

But the candidate $p$'s are not positive probabilities because the set of 
histories
\eqref{fourtwentyone} is not linearly positive and the inference cannot be
drawn.  In particular from \eqref{fourtwentytwo}
\begin{equation}%
\begin{split}
p(A, B|\Phi)=0, \quad  &p(A, \bar B|\Phi) = 1,  \\
p(\bar A, B|\Phi)=1, \quad  &p(\Bar A, \bar B|\Phi)=-1.
\end{split}
\label{fourtwentysix}
\end{equation}
The sum rules \eqref{fourtwentyfour} and \eqref{fourtwentyfive} are
exactly satisfied with $p(A|\Phi)=p(B|\Phi)=1$.

\subsection{A Single Particle}

The next example is more realistic.  We consider a single free
non-relativistic particle of mass $M$ moving in one dimension $x$. The
Hamiltonian is $H=p^2/2M$. For the initial state, we choose a Gaussian wave
packet of width $\sigma$ centered about the origin with zero expected value
for the momentum.  Specifically,
\begin{equation}
\Psi (x) = (2\pi\sigma^2)^{-\frac{1}{4}}\ e^{-x^2/4\sigma^2}
\label{fourtwentyseven}
\end{equation}   
We consider histories defined by exhaustive sets of position intervals at
the initial time $t=0$ and at a later time, $t=\tau$. For simplicity we
take the set of intervals $\{\Delta_\alpha\}$ ($\alpha$ an integer) to be
the same at both times.  A coarse-grained history is defined by the pair of
position intervals $\Delta_{\alpha_1}$ and $\Delta_{\alpha_2}$ 
the particle passes
through at the two times. The candidate probabilities for these histories
are denoted by $p(\alpha_2, \alpha_1)$ and constructed by
implementing \eqref{threeone} as follows:

Let $P_\alpha(t)$ denote the Heisenberg picture projection onto the range
$\Delta_\alpha$ of $x$ at time $t$.  The candidate probabilities for the
histories described above are [{\it cf.}~\eqref{twotwo}]
\begin{equation}
p (\alpha_2, \alpha_1) = {\rm Re}\ \left\langle\Psi\left|P_{\alpha_2}
(\tau)\, P_{\alpha_1} (0)\right|\Psi\right\rangle.
\label{fourtwentyeight}
\end{equation}
For purposes of computation, it is convenient to express the right-hand side
of \eqref{fourtwentyeight} in the Schr\"odinger picture. The connection is
made through
\begin{equation}
P_\alpha (t) = e^{iHt/\hbar} \hat P_\alpha e^{-iHt/\hbar}
\label{fourtwentynine}
\end{equation}
where $\hat P_\alpha$ denotes the Schr\"odinger picture projection onto the
range $\Delta_\alpha$ of $x$. The result for $p(\alpha_2,\alpha_1)$ is
\begin{equation}
{\rm Re} \int_{\Delta_{\alpha_2}} dx_2
\int_{\Delta_{\alpha_1}} dx_1 \Psi^* (x_2, \tau) K(x_2, x_1, \tau)
\Psi (x_1, 0).
\label{fourthirty}
\end{equation}
Here $\Psi (x_2, \tau)$ is the initial state \eqref{fourtwentyseven}
evolved to time $\tau$ and $K(x_2, x_1, \tau)$ is the free particle
propagator given by
\begin{equation}
\begin{split}
K(x_2, x_1, \tau) &\equiv \left\langle x_2\left|e^{-iH\tau/\hbar}\right|
x_1\right\rangle \\ &= \left(\frac{M}{2\pi i\hbar\tau}\right)^{\frac{1}{2}}
\exp \left[-\frac{M(x_2 - x_1)^2}{2i\hbar\tau}\right].
\end{split}
\label{fourthirtyone}
\end{equation}

The evolved Gaussian wave packet \eqref{fourtwentyseven} is
\begin{equation}
\begin{split}
\Psi (x, \tau) =& \frac{1}{(2\pi\sigma^2)^{\frac{1}{4}}} 
\ \left(1+ \frac{i\hbar\tau}{2\sigma^2 M}\right) \\ & \times\exp
\left[-\frac{x^2}{4\sigma^2}\ \left(1+ \frac{i\hbar\tau}{2\sigma^2
M}\right)^{-1}\right].
\end{split}
\label{fourthirtytwo}
\end{equation}
The expression for the candidate probabilities \eqref{fourthirty}
simplifies considerably if we restrict attention to times which are short
compared to the wave packet spreading time
\begin{equation}
\tau \ll \tau_{\rm spread} \equiv \frac{2\sigma^2 M}{\hbar} \approx  
(2\times 10^{27} {\rm s})\ \left(\frac{\sigma}{1\ {\rm cm}}\right)^2
\left(\frac{M}{1\ {\rm gm}}\right).
\label{fourthirtythree}
\end{equation}
There are many interesting situations in which this is a realistic
assumption.  In this approximation
\begin{widetext}
\begin{equation}
p (\alpha_2, \alpha_1) = \left(\frac{1}{2\pi
\sigma^2}\right)^{\frac{1}{2}} \left(\frac{M}{2\pi \hbar
\tau}\right)^{\frac{1}{2}} \int_{\Delta_{\alpha_2}} dx_2
\int_{\Delta_{\alpha_1}} dx_1 \exp \left[-\frac{\left(x^2_2 +
x^2_1\right)}{4\sigma^2}\right] \cos \left[\frac{M(x_2-x_1)^2}{2\hbar\tau}
- \frac{\pi}{4}\right].
\label{fourthirtyfour}
\end{equation} 
\end{widetext}
We now discuss the behavior of these candidate probabilities for a simple
coarse graining of this set of histories.

The probabilities for the coarse graining we consider answer the question:
Does the particle remain at the origin or does it move elsewhere?
Specifically, we consider the set consisting of just two histories: The
history $L$ in which the particle is localized
in an interval $\Delta$ centered on
the origin at both times $t_1$ and $t_2$, and the history $\bar L$ in 
which it is not.

The candidate probability $p_L (\Delta)$ for the particle to be
localized at both times is given by \eqref{fourthirtyfour} with
$\Delta_{\alpha_1} = \Delta_{\alpha_2} \equiv \Delta$. The candidate
probability $p_{\bar L} (\Delta)$ that is not so localized, can be found
from this and
\begin{equation}
p_L (\Delta) + p_{\bar L} (\Delta)=1.
\label{fourthirtyfive}
\end{equation}

Equation \eqref{fourthirtyfour} can be organized to give a tractable
expression for $p_L(\Delta)$ by defining new integration variables
\begin{equation}
X=(x_1 + x_2)/2\ ,\quad \xi=x_2-x_1,
\label{fourthirtysix}
\end{equation}
and introducing a characteristic wavelength of oscillation
\begin{equation}
\lambda \equiv 
\left[4\pi\left(\frac{\hbar\tau}{M}\right)\right]^{\frac{1}{2}}
= 1.1 \times 10^{-13} \left(\frac{\tau}{1\ s}\right)^{\frac{1}{2}}
\left(\frac{1\ {\rm g}}{M}\right)^{\frac{1}{2}} {\rm cm}.
\label{fourthirtyseven}
\end{equation}
Then, after some algebra
\begin{subequations}
\label{fourthirtyeight}
\begin{equation}
p_L (\Delta) =   \sqrt{\frac{2}{\pi\sigma^2}} \int^\Delta_0 dX
\ e^{-X^2/2\sigma^2} J(\Delta-X)
\label{fourthirtyeighta}
\end{equation}
where
\begin{equation}
\begin{split}
J(z)  &\equiv  \frac{2^{\frac{3}{2}}}{\lambda} \int^z_0 d\xi
\ e^{-\xi^2/8\sigma^2} \cos \left[2\pi
\left(\frac{\xi}{\lambda}\right)^2 - \frac{\pi}{4}\right]
\\
 &=   {\rm Re}\ \left\{{\rm Erf} \left[\sqrt{\pi} (1-i)
\frac{z}{\lambda}\right]\right\}.
\end{split}
\label{fourthirtyeightb}
\end{equation}
\end{subequations}
These expressions are quoted in the approximation $\lambda \ll \sigma$
typically valid for large particles as \eqref{fourthirtyseven} shows. The
general ones are not much more complicated.

\begin{figure}[t]
\centerline{\epsfysize=2in \epsfbox{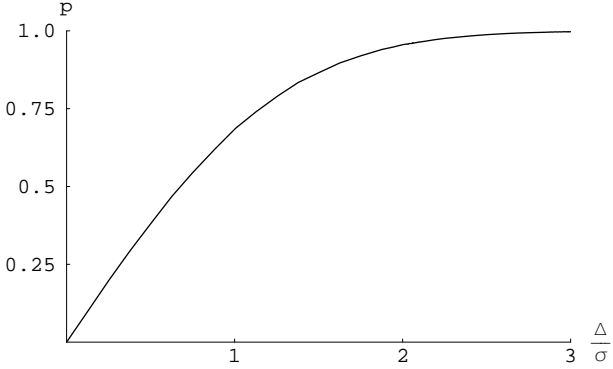}}
\caption{Candidate probabilities for a single free particle in a stationary
Gaussian wave packet to remain in a position interval $\Delta$ after a time
short compared to the wave-packet spreading time. The plot shows the
candidate probability $p_L (\Delta)$ defined by \eqref{fourthirtyeighta}
for $\lambda/\sigma=10^{-2}$.
This is positive over the whole of the range $\Delta$ indicating that the
set consisting of the history in which the particle remains localized and
the history where does not so remain is linearly positive.}
\end{figure}

As $\Delta$ becomes large, $p_L(\Delta)$ must approach unity and
\eqref{fourthirtyeight} exhibits this explicitly $({\rm Erf} (\infty)=1)$. 
But also
from \eqref{fourtwentyeight}, and the general result \eqref{threetwo}, 
$p_L(\Delta)\leq 1$. For sufficiently large $\Delta$ this set of histories
is linearly positive.  Analysis of the situation for $\Delta \ltwid \sigma$
requires numerical integration of \eqref{fourthirtyeight} but the following
result emerges: At least for $\lambda/\sigma \sim 1/100$, 
$p_L(\Delta)$ is positive over the whole range of $\Delta$ from 0 to
$\infty$. Figure 4 shows $p_L(\Delta)$ for this case.
Taken together with $p_L (\Delta) \leq 1$, this means that this
set is always positive.  The quantities $p_L(\Delta)$ and $p_{\bar L}
(\Delta)$ are genuine probabilities.

Although linear positive, this set of histories is far from decoherent.
With $C_L \equiv P_\Delta (\tau) P_\Delta (0)$ the off-diagonal elements of
the decoherence functional are
\begin{equation}
\begin{split}
D(L, \bar L) & \equiv \langle\Psi|C^\dagger_L C_{\bar
L}|\Psi\rangle = \langle\Psi|C^\dagger_L
(I-C_L)|\Psi\rangle \\ & = \langle \Psi|P_\Delta (0) P_\Delta (\tau)
P_{\bar \Delta} (0)|\Psi\rangle
\end{split}
\label{fourthirtynine}
\end{equation}
where $P_{\bar \Delta}$ is the projection on the range outside $\Delta$ at
$t=0$.
We can think of \eqref{fourthirtynine}
as the overlap of two states $|\Psi_L\rangle\equiv P_\Delta
(\tau) P_\Delta (0)|\Psi\rangle$ and $|\Psi_{\bar L}\rangle\equiv
P_\Delta (\tau) P_{\bar
\Delta} (0)|\Psi\rangle$. When $\Delta \gg \sigma$, $|\Psi_{\bar L} \rangle$ 
will have negligible length because $P_{\bar\Delta} (0)|\Psi\rangle$ 
is negligible.
Consider therefore $\Delta \ll \sigma$. The wave function of $P_{\bar
\Delta}(0)|\Psi\rangle$ consists of those parts of the initial packet
\eqref{fourtwentyseven} outside of $\Delta$.  Because of the sharp interior
edges, this evolves rapidly to fill in the center around $x=0$. After
projection of $P_\Delta (\tau)$ there is significant overlap with the wave
function of $|\Psi_L\rangle$ of order $(\Delta/\sigma)$ and thus absence of
decoherence. 

In general we expect the action of an environment to be necessary to carry
away the phases between such alternatives and make them decoherent ({\it
e.g.} \cite{GH93a}). Despite the absence of an environment
and its consequent decoherence this set of
histories is linearly positive in the approximations considered.

\subsection{A Spacetime Alternative}

The familiar example of a coarse-grained history is a sequence of events at
a series of definite times. But decoherent histories quantum theory permits
more general coarse grainings that extend continuously over ranges of time
\cite{Har91b,YT91}. 
Such spacetime coarse grainings may provide more realistic models of
measurement processes that extend over time. Analogous coarse grainings may
be essential for a quantum theory of gravity where there is no fixed notion
of time ({\it e.g.} \cite{Har95c}). 

To illustrate the idea of a spacetime coarse graining, consider the motion
of a single free particle of mass $M$ moving in one dimension $x$ with
Hamiltonian $H=p^2/2M$ over a range of times from 0 to $T$. The set of
fine-grained histories of the particle's motion consists of the paths $x(t)$ on the
interval $[0, T]$. Sets of alternative coarse-grained histories are defined
by partitions of these fine-grained histories into mutually exclusive
classes $c_\alpha, \alpha = 1, 2, \cdots$. Each class is a
coarse-grained history.

The class operators $C_\alpha$ for these coarse-grained histories can be
constructed from sums over their constituent fine-grained histories. To see
how, first consider a partition of the paths by an exhaustive set of
position intervals $\{\Delta_\alpha\}$ at a series of times $t_1< \cdots<
t_n$. The class operator in the Heisenberg picture is $C_\alpha =
P_{\alpha_n}(t_n) \cdots P_{\alpha_1} (t_1)$ where $P_\alpha (t)$ is the
projection on the range $\Delta_\alpha$ at time $t$. Matrix elements of the
$C_\alpha$ in the Heisenberg picture (HP) can be transformed into transition
amplitudes in Schr\"odinger picture (SP), and these transition amplitudes
can be expressed as path integrals. The result is
\begin{equation}
\begin{split}
\langle\Phi|C_\alpha|\Psi\rangle_{\rm HP} &= \langle\Phi (T)|\hat
C_\alpha|\Psi (0)\rangle_{\rm SP}, \\&= \int_{c_\alpha} \delta x\ \Phi^* (x_T,
T)\ e^{iS[x(t)]/\hbar}\ \Psi (x_0, 0).
\end{split}
\label{fourforty}
\end{equation}
Here,
\begin{equation}
\hat C_\alpha \equiv e^{-iHT/\hbar} C_\alpha,
\label{fourfortyone}
\end{equation}
and the path integral is over all paths $x(t)$ in the class $c_\alpha$
including an integral over the end points $x_0$ and $x_T$. (More details on
establishing this relation can be found in \cite{Har91b} and \cite{Cav86}.)
The candidate probabilities for the coarse-grained histories are, from
\eqref{threeone}
\begin{equation}
p (\alpha) = {\rm Re}\ \langle\Psi(T)|\hat C_\alpha|\Psi(0)\rangle_{\rm SP}.
\label{fourfortytwo}
\end{equation} 
From now on we drop the SP's and HP's and rely on context to distinguish
the two pictures.

The derivation sketched above of the connection \eqref{fourforty} between
matrix elements of class operators and path integrals over classes was for
coarse grainings by ranges of position at a sequence of times. But the
result motivates using path integrals to {\it define} class operators for
arbitrary partitions of the fine-grained histories $x(t)$ into mutually 
exclusive classes including partitions defining alternatives extending over
a range of time. Eq.~\eqref{fourfortytwo} gives candidate probabilities for
these spacetime coarse grainings.

A simple model illustrates the idea \cite{Har91b,MH96}. Partition all paths
$x(t)$ on the interval $[0,T]$ into the two classes
\begin{itemize}

\item[] $R$:  paths $x(t)$ that {\it always} remain in the region \\ $x>0$ between times 0
and $T$.
\item[] $\bar R$: paths $x(t)$ that {\it sometimes} are in the region \\ $x<0$
between times 0 and $T$.
\end{itemize}
Evidently, these classes are exhaustive and mutually exclusive.  $R$ here
means ``right'', $x>0$, and $p_R$ computed from \eqref{fourfortytwo} is the
probability that the particle remains at $x>0$ for the whole time
interval and never crosses into $x<0$.

A sum over all paths of the form \eqref{fourforty} that is restricted
to $x>0$ is the same as an unrestricted sum in the presence of an infinite
potential barrier at values of $x<0$ \cite{Har91b}. Let $H_R$ denote the
Hamiltonian of the particle including this barrier. The branch state vector
for the class $R$ can be written
\begin{equation}
|\Psi_R(T)\rangle = P_R e^{-iH_R T/\hbar} P_R |\Psi (0)\rangle
\label{fourfortythree}
\end{equation}
where $P_R$ is the projection onto $x>0$. The candidate probability $p_R$
is [{\it cf.}~\eqref{fourfortytwo}]
\begin{equation}
p_R = {\rm Re}\ \langle\Psi(T)|\Psi_R(T)\rangle.
\label{fourfortyfour}
\end{equation}
It's not necessary to derive a separate expression for $p_{\bar R}$ in this
simple example.  From \eqref{oneone}, \eqref{fourthirtynine} and
\eqref{fourforty} it follows that
\begin{equation}
p_R + p_{\bar R} = 1
\label{fourfortyfive}
\end{equation}
(whether or not the set is linearly positive.) Calculation of $p_R$ will
therefore determine $p_{\bar R}$.

For explicit calculation it is convenient to rewrite \eqref{fourfortyfour}
in terms of wave functions. Let $\Psi(x,0)$ denote the wave function
 of the initial state and $\Psi_U (x,T)$ its
unrestricted evolution under $H$. Let $\Phi(x,0) \equiv P_R \Psi (x,0)$, and
denote by $\Phi_R(x, T)$ and $\Phi_U(x,T)$ its evolution under $H_R$ and
$H$ respectively. Evolution in the presence of an infinite barrier is
especially simple, and
\begin{equation}  
\Phi_R (x,T) = \Phi_U (x,T) - \Phi_U (-x,T)\ ,\ x>0.
\label{fourfortysix}
\end{equation}
Thus, we can write
\begin{equation}
\begin{split}
p_R&= {\rm Re}\int\limits^\infty_0 dx\ \Psi^*_U (x,T)\, \Phi_R (x,T) \\& =
{\rm Re}\int\limits^\infty_0 dx\, \Psi^*_U (x,T)\ \left[\Phi_U (x,T) - \Phi_U
(-x,T)\right].
\end{split}
\label{fourfortyseven}
\end{equation}
This form allows both quantitative computation and qualitative discussion.
Since
\begin{equation}
\||\Phi_R (T)\rangle\| = \|| \Phi_R (0)\rangle\| = \||P_R\Psi(0)\rangle\|,
\label{fourfortyeight}
\end{equation}
an immediate consequence of \eqref{fourfortyseven} is
\begin{equation}
p_R \leq \|P_R|\Psi_U (T)\rangle\|\ \|P_R|\Psi(0)\rangle\|.
\label{fourfortynine}
\end{equation}
This shows $p_R \leq 1$ and $p_{\bar R} \geq 0$.

We now evaluate \eqref{fourfortyseven} for a very simple initial wave
function $\Psi(x,0)$. We consider a Gaussian wave packet of width $\sigma$
centered about an initial position $X_0>0$ and moving toward negative $x$
with a (negative) momentum $K_0$. Specifically, assume
\begin{equation}
\Psi(x,0) = (2\pi\sigma^2)^{-\frac{1}{4}}\ e^{iK_0x}
e^{-(x-X_0)^2/(4\sigma^2)}.
\label{fourfifty}
\end{equation}
To keep expressions simple, we use units when $\hbar=1$ as we will for the
remainder of this section.

The following approximations make the evaluation of the integral
\eqref{fourfortyseven} for $p_R$ straightforward. 
\begin{enumerate}
\item We assume that 
\begin{equation}
X_0 \gg \sigma
\label{fourfiftyone}
\end{equation}
so that the initial wave function is negligible for $x<0$, {\it i.e.} 
\begin{equation}
\Phi (x,0) \equiv P_R \Psi (x,0) \approx \Psi (x,0).
\label{fourfiftytwo}
\end{equation}

\item We assume that the time $T$ is short compared to the time over which
the wave packet spreads significantly, specifically
\begin{equation}
T\ll 2\sigma^2 M.
\label{fourfiftythree}
\end{equation}
\end{enumerate}
With these two approximations, the shape of the wave packet remains
approximately
unchanged under evolution over the time $T$. Only the center shifts 
to the value
\begin{equation}
X=X_0 + K_0 T/M.
\label{fourfiftyfour}
\end{equation}
Specifically,
\begin{equation}
\begin{split}
\Phi_U (x,T) &\approx \Psi_U (x, T) \\&\approx (2\pi\sigma^2)^{-\frac{1}{4}}
e^{iK_0x} e^{-(x-X)^2/4\sigma^2}\ e^{-i(K_0^2/2M)T}.
\end{split}
\label{fourfiftyfive}
\end{equation}
The integral \eqref{fourfortyseven} for $p_R$ is then
\begin{equation}
\begin{split}
p_R&\approx (2\pi\sigma^2)^{-\frac{1}{2}} \int\limits^\infty_0 dx
\ e^{-(x-X)^2/2\sigma^2} \\&\times\left[1-e^{-xX/\sigma^2} \cos (2K_0 x)\right]
\end{split}
\label{fourfiftysix}
\end{equation}
which is straightforward to evaluate numerically.

Assumption \eqref{fourfiftyone} means that $K_0$ must be large for the
center of the wave packet to reach the neighborhood of $x=0$ in the times
limited by \eqref{fourfiftythree}. Specifically, $K_0\sigma\gg 1$. Otherwise,
the wave packet remains to the right of $x=0$ and $p_R\approx 1$.

\begin{figure}[t]
\centerline{\epsfysize=2in \epsfbox{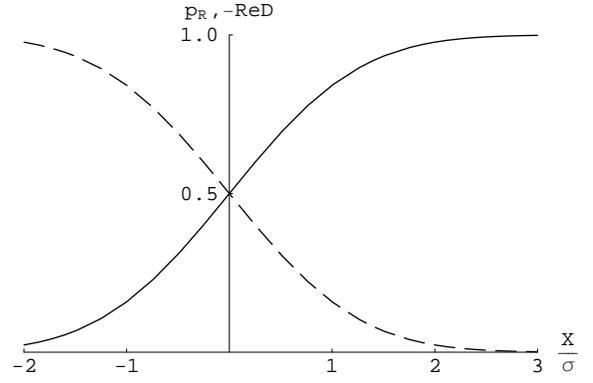}}
\caption{Candidate probabilities for a spacetime coarse graining. A free
particle moving in one dimension is initially in a Gaussian wave packet of
width $\sigma$ traveling toward negative $x$ with a momentum $K_0 = -
20/\sigma$ in $\hbar=1$ units. The solid curve shows the candidate
probability $p_R$ for
the particle to remain always at positive $x$ over a time $T$ short
compared to the wave packet spreading time. This is plotted as a function
of the final position of the packet's center $X$ that is connected to the
initial position $X_0$ and $T$ by \eqref{fourfiftyfour}.
The probability is positive over the
entire range computed showing that the set consisting of the history $(R)$,
where the particle remains at positive $x$, and the history $(\bar R)$,
where it does not, is linearly positive. The dashed curve shows the real
part of the overlap $\langle \Psi_R|\Psi_{\bar R}\rangle$ which must vanish
for this set of histories to be medium decoherent.  Evidently there is a
significant range of linearly positive sets which are not medium
decoherent.}
\end{figure}

Figure 5 shows $p_R$ from \eqref{fourfiftysix} as a function of the
position of the center $X$ at time $T$ for $K_0= -20/\sigma$. Only values of
$X\gtwid -2$ are shown for which $X_0 \gtwid 3\sigma$ in a time $T$ about 1/10
of the spreading time in \eqref{fourfiftythree}. For smaller values of $X$,
the approximation \eqref{fourfiftyfive} would be inaccurate.

For large values of $X$, the wave packet is localized some place in $X>0$
over the whole of the time interval. We might therefore expect $p_R\approx
1$, and it is.  For smaller $X$, there is a significant probability of
crossing $X=0$ at least once in the time interval 0 to $T$. The important
point is that, over the whole of the calculated range, $p_R$ lies between 0
and 1. Likewise for $p_{\bar R}$ from \eqref{fourfortythree}. The set of
alternatives is thus linearly positive with genuine probabilities.  We now
turn to the question of whether this set of alternatives decoheres.

The set of coarse-grained histories $R$ and $\bar R$ decoheres when [{\it
cf.}~\eqref{onefour}]
\begin{equation}
D(R, \bar R) \equiv \langle \Psi_R (T) |\Psi_{\bar R} (T)\rangle \approx 0.
\label{fourfiftyseven}
\end{equation}
The branch state vector $|\Psi_R\rangle$ was exhibited in
\eqref{fourfortythree}.
Relation \eqref{oneone} and \eqref{fourfortyone} imply that
\begin{equation}
|\Psi_{\bar R} (T)\rangle = |\Psi (T)\rangle - |\Psi_R (T)\rangle.
\label{fourfiftyeight}
\end{equation}
In terms of wave functions
\begin{equation}
D(R, \bar R) = \int\limits^\infty_0 dx\ \Phi^*_R (x, T)\ \left[\Psi_U (x,T)
- \Phi_R (x,T)\right]
\label{fourfiftynine}
\end{equation}
where $\Phi_R (x,T)$ can be taken to be given by \eqref{fourfortysix}. In the
approximation where \eqref{fourfiftytwo} holds, this is
\begin{equation}
D (R, \bar R) \approx \int\limits^\infty_0 dx\ \left[\Psi^*_U (x,T) -
\Psi^*_U (-x, T)\right]\, \Psi_U (-x, T).
\label{foursixty}
\end{equation}
Figure 5 shows ${\rm Re}\ D(R, \bar R)$ calculated from \eqref{foursixty}
plotted against $X$. For large positive $X$, the particle remains on the 
right. There is thus essentially only one history with significant
probability, $|\Psi_R\rangle \approx |\Psi\rangle$, $|\Psi_{\bar R}\rangle
\approx 0$ and decoherence is automatic. For large negative $X$, almost
all of the wave packet $\Psi_U(x,T)$ will have crossed $x=0$. That is
outside the range where \eqref{foursixty} is necessarily a good
approximation, but the first term will vanish and $\langle\Psi_R |\Psi_{\bar
R}\rangle \approx -1$. That is far from decoherence.

Figure 5 shows that over a wide range of situations for which this set of
histories is linearly positive, it is not decoherent. That is consistent
with linear positivity being a weaker condition for probabilities.   

\section{Virtual Probabilities}

What about sets of histories that are not linearly positive?  Can the
candidate probabilities $p (\alpha)$ defined by \eqref{threeone} be put 
to use even if
they are outside the range 0 to 1? This section will show that they can.
In particular, $p (\alpha)$ outside the range 0 to 1 can be employed
as intermediate steps in the calculation of probabilities that do lie
between 0 and 1. That is because probability sum rules like \eqref{twofourb} 
are satisfied by the $p(\alpha)$ defined by \eqref{twoone} whether or not they
lie between 0 and 1 as a consequence of \eqref{twothree}. For this reason,
if some $p (\alpha)$ in a set are outside the range 0 to 1, we
say the set has {\it virtual probabilities}.  When a contrasting term is
needed for the case when all the $p (\alpha)$ are between 0 or 1, we say the set
has {\it real probabilities}.\footnote{To maintain the usual
contrast between virtual and real, we accept the risk that ``real'' can be
taken to be a contrast with ``imaginary''.}
``Real'' in this context is
thus a synonym for linearly positive.
Extending sets of probabilities to include virtual
values provides a simple and unified approach to the quantum mechanics of
closed systems as we shall describe in Section VI. 
Anticipating this extension, in this Section we refer to
the candidate probabilities $p(\alpha)$ defined by \eqref{threeone} as 
``probabilities''
whether or not they are real or virtual.

To understand what virtual probabilities might mean, first consider what
they cannot mean. Evidently, virtual probabilities cannot coincide with the
relative frequencies of repeated events such as the probabilities of the
outcomes of identical measurements on an ensemble of identical subsystems.
More generally, alternatives with virtual probabilities cannot be exactly
recorded or measured. The exact correlation between alternative values of
records represented by {\it projections} $\{R_\beta\}$ and  histories
represented by class operators $\{C_\alpha\}$ would mean [{\it cf.}
\eqref{threeseven}] 
\begin{equation}
p(\beta, \alpha) \equiv {\rm Re} \left\langle\Psi\left|R_\beta
C_\alpha\right|\Psi\right\rangle = \delta_{\alpha\beta} p(\alpha).
\label{fiveone}
\end{equation}
Sum both sides of this relation over $\beta$ using $\sum_\beta R_\beta=I$.
Sum both sides over $\alpha$ using $\sum_\alpha C_\alpha =I$. The resulting
relations imply the identity
\begin{equation}
p (\alpha) \equiv {\rm Re}\ \left\langle\Psi
\left|C_\alpha\right|\Psi\right\rangle = {\rm Re}\, \left\langle \Psi \left|
R_\alpha\right|\Psi\right\rangle
\label{fivetwo}
\end{equation}
showing that the
probabilities of the records are the same as the probabilities of the
histories.  Since the $\{R_\alpha\}$ are projections, their probabilities
must be between 0 and 1.  Sets of histories with virtual
probabilities therefore cannot be exactly recorded.  
Since recording outcomes is usually
taken to be an essential part of a measurement process\footnote{See
{\it e.g.} \cite{Har91a}}, we
can say that sets of histories with virtual probabilities cannot 
describe the outcomes of measurements. Extending the notion of probability
to include virtual values thus does not risk assigning a virtual value to
the probability of anything measured or recorded exactly. Of course, the
records of realistic measurements are typically correlated with the
measured history, not exactly, but rather to an excellent approximation.

Virtual
probabilities can in principle be employed in the calculation of real ones
as the following example illustrates.
Consider the
probabilities $p(\gamma|pd)$ of some future alternatives $\{\gamma\}$
given present data $pd$. It may be more efficient to calculate these
probabilities by first determining the probabilities $p(\beta|pd)$
of alternatives $\{\beta\}$ in the past and then with these, and with
present data, calculate the probabilities of the future alternatives
$\{\gamma\}$. More concretely, it may be efficient to calculate 
$p(\gamma|pd)$ from the relation
\begin{equation}
p (\gamma|pd) = \sum\nolimits_\beta 
p(\gamma|pd, \beta)\, p(\beta|pd).
\label{fivethree}
\end{equation}
For example, if a current manuscript $(pd)$ records that Caesar invaded
Britain in 55 BC, we might predict that other manuscripts yet to be
discovered $(\{\gamma\})$ would record the same date. However, the most
direct route to that prediction would be to first infer that Caesar did
invade Britain in 55 BC (one of the alternatives $\beta$) and from that
calculate the probabilities of what other manuscripts might say.  We
reconstruct the past to help explain the future.

Our purpose here is not to discuss the utility of reconstructing the 
past\footnote{For that see \cite{Har98b}}.
Rather, it is to point out that \eqref{fivethree} is a
consequence of \eqref{twoone} and \eqref{threeone} which hold whether or not the probabilities
$p(\beta|pd)$ are between 0 and 1. In principle, virtual 
$p(\beta|pd)$ could be used in the intermediate step in
\eqref{fivethree} provided the result $p(\gamma|pd)$ were real
probabilities.

The three-box example discussed in Section IV.C provides a simple, if
artificial, example.  Suppose we are interested in predicting probabilities
for whether the particle will be either in box $A$, or 
else not in $A$ ({\it i.e.}, in box $B$ or $C$) 
at some future time given $\Phi$. 
We denote these alternatives by $A^f$ and $ \bar A^f$. The
probability $p(\bar A^f|\Phi)$ for instance is given by
\begin{equation}
p(\bar A^f|\Phi) = 
\frac{p(\bar A^f, \Phi)}{p(\Phi)}
 = \frac{{\rm Re}\ \left\langle\Psi\left|\left(P_B +
P_C\right)P_\Phi\right|\Psi\right\rangle}{{\rm Re}\ \langle\Psi|P_\Phi|
\Psi\rangle}
\label{fivefour} 
\end{equation}
From \eqref{foursixteen} and \eqref{fourseventeen} we find,
\begin{equation}
p \left(A^f|\Phi\right)=1\ ,\quad  p(\bar A^f| \Phi)=0
\label{fivefive}
\end{equation}
--- a set of real probabilities.

But we could calculate these probabilities by first calculating the $p$'s
for the alternatives $A^p, B^p, C^p$ that the particle was in box
$A$, $B$, or $C$ in the past given $\Phi$ and then using \eqref{fivethree}
to calculate the future probabilities for $A^f$ and $\bar A^f$. 
The probabilities for the past alternatives $A^p, B^p, C^p$ given $\Phi$
can be negative, for example
\begin{equation}
p\left(C^p| \Phi\right) = \frac{p(\Phi, C^p)}{p(\Phi)} = 
\frac{{\rm Re}\ \left\langle\Psi \left|P_\Phi P_C \right|\Psi 
\right\rangle}{{\rm Re}\ \left\langle\Psi\left|P_\Phi\right|\Psi\right\rangle} = -1.
\label{fivesix}
\end{equation}
Similarly, $p(A^p|\Phi)=p(B^p|\Phi)=1$. Despite this the
reader may easily verify that employing \eqref{fivethree} with $\{\gamma\} =
\{A^f, \bar A^f\}, \{\beta\} = \{A^p, B^p, C^p\}$ and $pd=\Phi$
gives the correct answer displayed in \eqref{fivefive}.

Feynman \cite{Fey87} explored the uses of negative probabilities in
intermediate steps in a variety of circumstances in physics. These included
the probabilities for position and momentum defined by the Wigner 
distribution, for
the emission of virtual non-transverse photons in electrodynamics, in 
two-state systems, and in the two-slit experiment. He concluded that extending
the notion of probability to negative values was useful provided these
negative values are interpreted to mean that the situation is
``unattainable or unverifiable''. We take the same viewpoint here.

The quantum mechanics of a closed system then can be formulated as follows:
{\it Assign probabilities to all sets of alternative coarse-grained histories by
$p(\alpha)= {\rm Re}\ \langle\Psi|C_\alpha|\Psi\rangle$. The probabilities
$p(\alpha)$ may be real or virtual. Virtual probabilities can be
employed in the calculation of real ones. Probabilities predicted for
exactly recorded histories are always real}. 
This is a simple and general
formulation whose utility we describe in the next section. Linear
positivity and medium decoherence are special cases.  

\section{Comparing Decoherence Conditions}

Four decoherence conditions restricting the sets of alternative
coarse-grained histories to which quantum mechanics predicts probabilities
have been considered in this paper.  In order of increased restriction
(roughly inverse to the order presented) they are:
\begin{itemize}

\item {\sl Extended Probabilities (EP)}\\
Assign probabilities to all sets of alternative coarse-grained histories
according to 
\begin{equation}
p (\alpha) = {\rm Re}\ \langle\Psi|C_\alpha|\Psi\rangle
\label{sixone}
\end{equation}
whether they are real (between 0 and 1) or virtual.  This is no restriction
at all. Calculate the probabilities of exactly recorded alternatives, or
the frequencies of repeated independent alternatives, secure in the
knowledge that these will be between 0 and 1.

\item {\sl Linear Positivity (LP)}\\
Assign probabilities only to sets of alternative coarse-grained histories
for which all candidate probabilities in the set given by \eqref{sixone} 
lie in the
range 0 to 1 of real probabilities.

\item{\sl Real Linear Positivity (RLP)} \\ 
 Assign probabilities only to sets of alternative
coarse grained histories which are linearly positive and for which 
\begin{equation}
{\rm Im }[\langle \Psi |C_\alpha | \Psi \rangle]  \approx 0, \quad  \text{all $\alpha$,}
\label{sixonea}
\end{equation}
either exactly or approximately well beyond the standard with which the resulting
probabilities are used. 

\item {\sl Medium Decoherence (MD)}\\
Assign probabilities only to sets of alternative coarse-grained histories
for which the branch state vectors for all histories in the set are
mutually orthogonal
\begin{equation}
\left\langle\Psi_\alpha |\Psi_{\alpha^\prime}\right\rangle \approx 0\ ,
\ \alpha \not= \alpha^\prime
\label{sixtwo}
\end{equation}
either exactly, or to an approximation well beyond the standard with which the
resulting probabilities are used. Calculate probabilities either by 
\eqref{sixone} or by
\begin{equation}
p (\alpha) = \left\| |\Psi_\alpha\rangle\right\|^2 = \left\| C_\alpha
|\Psi\rangle\right\|^2
\label{sixthree}
\end{equation}
which are equivalent if medium decoherence is exact. Probabilities defined by
\eqref{sixthree} necessarily lie between 0 and 1.

\end{itemize}
This concluding section compares these different decoherence conditions.

As information gathering and utilizing systems (IGUS's), we employ almost
exclusively coarse grainings of the usual quasiclassical realm. By the usual
quasiclassical realm, we mean roughly histories of coarse-grained alternatives
defined by ranges of values of averages of densities of approximately
conserved quantities (such as energy, momentum, etc.) over suitable
volumes. With the initial condition and Hamiltonian of our universe, the
volumes can be chosen large enough that the histories are medium
decoherent and yet small enough to supply a reasonably fine-grained
description of the universe over a wide range of time and distance scale.
Individual histories of this realm exhibit patterns of correlations in time
governed by effective classical equations of motion interrupted by
frequent small quantum fluctuations and occasional major ones.

For practical purposes it might be possible to restrict the predictions of
quantum theory to the usual quasiclassical realm. Indeed, in some loose
sense, this was the view of founders of the subject such as Bohr and
Heisenberg. But it so far has been difficult to give the usual
quasiclassical realm a precise definition despite the considerable steps
that have been taken in that direction \cite{fnnine}.
For reasons of generality,
convenience, and completeness it has proved useful to formulate quantum
theory with weaker, less anthropocentric, but more precise conditions that
allow many more sets of histories that are nothing like the usual
quasiclassical realm. The three conditions discussed in this paper are
examples.

Moving between EP, LP, RLP, and MD in the  of  increasing restriction is
moving toward the usual quasiclassical realm. For example, the virtual
probabilities of EP for non-equal values of a quantity commuting
with the Hamiltonian at two different times may not be zero if other
alternatives intervene as in \eqref{threenine}. But they are zero under
 the more restrictive LP
conditions as the discussion in Section III.C shows.

The conditions RLP and MD are consistent with the usual notions of statistical
independence of identical subsystems; EP and LP are not. 

The conditions EP,  LP and RLP are time-neutral, but the more restrictive MD incorporates
an arrow of time as discussed in Section III.C. This arrow allows
individual histories to be described as a narrative in which one event
is followed by another, then another, etc. Quantum theory then can be
formulated in terms of
an evolving state, and prediction distinguished from retrodiction. For
example, consider a medium decoherent set of histories defined by
alternatives $\{P^1_{\alpha_1} (t_1)\}$, $\{P^2_{\alpha_2}(t_2)\}, \cdots
\{P^n_{\alpha_n} (t_n)\}$
at a series of times $t_1 < t_2 < \cdots < t_n$. At any intermediate time
$t_k$, we can calculate the conditional probability $p(\alpha_n, \cdots,
\alpha_{k+1}|\alpha_k, \cdots, \alpha_1)$ for future alternatives
$\alpha_{k+1}, \cdots, \alpha_n$ given that $\alpha_1, \cdots, \alpha_k$
have already happened by
\begin{equation}
\begin{split}
&p\left(\alpha_n,  \cdots, \alpha_{k+1}|\alpha_k, \cdots,
\alpha_k\right) \\ &= \left\|P^n_{\alpha_n} (t_n) \cdots P^{k+1}_{\alpha_{k+1}}
(t_{k+1}) |\Psi_{\alpha_k \cdots \alpha_1}\rangle \right\|^2
\end{split}
\label{sixfour}
\end{equation}
where
\begin{equation}
\bigl|\Psi_{\alpha_k \cdots \alpha_1}\bigr\rangle = \frac{P^k_{\alpha_k}
(t_k) \cdots P^1_{\alpha_1}
(t_1)|\Psi\rangle}{\left\|P^k_{\alpha_k}(t_k) \cdots P^1_{\alpha_1}
(t_1)|\Psi\rangle\right\|}.
\label{sixfive}
\end{equation}
Events to the future of $t_k$
can be predicted just from the state $|\Psi_{\alpha_k \cdots
\alpha_1}\rangle$ representing the present. In the Heisenberg picture used
here, that state is constant in time except when interrupted by the action
of projections (``reductions'') representing the alternatives.

The past cannot be retrodicted just from a state in the present, but
requires in addition the state $|\Psi\rangle$. A similar statement holds
for the future in the context of LP or EP in general 
(see, {\it e.g.}~\cite{Har98b,GH90a}).
It could not be otherwise since these formulations are
time-neutral.    

Histories of the usual quasiclassical realm are medium decoherent 
because of physical mechanisms which dissipate phases between branches
\cite{JZ85, CL83, Zur84} that result from the  interaction between the
variables followed in the usual quasiclassical coarse graining and ones
ignored constituting an environment or bath in simple models. These
interactions create records of the histories \cite{Hal99,Hal03} 
and the existence
of records is a general characterization of medium decoherence as we saw in
Section III.A. The restriction to medium decoherence is thus consistent
with the usual quasiclassical realm.

Imposing restrictive decoherence conditions like medium and strong
decoherence \cite{GH95} are useful first steps in defining classicality.
But the weaker conditions (LP) and (EP) also have their uses.

An example is approximate medium decoherence.  As mentioned above, the 
branch state vectors of the usual quasiclassical realm are not expected to
be exactly orthogonal, but only to an approximation good well beyond
the standard to which the resulting  approximate probabilities can be used.
Some are uneasy about basing a fundamental formulation of quantum
mechanics on {\it approximate} medium decoherence \cite{DK96}. Is there a
formulation of the quantum mechanics of closed systems free from any
approximate notion of decoherence for realistic coarse grainings? There is;
consider the following possibility: Regard linear positivity (or EP) as the
fundamental rule determining the probabilities of histories in quantum
mechanics. The resulting probabilities satisfy the sum rules exactly as
discussed in Section III.B.  Medium decoherence then becomes an approximate
notion, useful in characterizing classicality, giving through
\eqref{threefive} an approximation to the fundamental probabilities
\eqref{sixone}. The degree of approximation can be calculated from
\eqref{threesix}. No practical calculation of the probabilities of the
quasiclassical realm or its coarse grainings is likely to be affected by
adopting this viewpoint.  A consequence, however, is that many more sets of
coarse-grained alternative histories nothing like the quasiclassical realm
are incorporated into the theory beyond the already large number allowed by
approximate medium decoherence. A further consequence is that connection
between probabilities and the frequencies of even imaginary ensembles of
independent subsystems is lost in general as discussed in Section
III.D. Precision is achieved at the
consequence of extending the complementary descriptions of the world and
possibly at the expense of any general frequency interpretation of the resulting probabilities.

A more important reason for considering linear positivity or extended
probabilities as a fundamental bases for quantum theory lies in
their potential for further generalization. The quantum theory of closed
systems summarized in Section II may need to be further generalized to
incorporate quantum gravity.  That is because the framework in Section II
relies on a notion of time supplied by a fixed background spacetime
geometry. But, in general relativity, spacetime geometry is a dynamical
variable that generally will fluctuate and be without a definite value in a
quantum theory of gravity. A generalization of the usual quantum framework
is required.  

Generalized quantum theory \cite{Har91a,GH90a,Ish94,IL94,Har95c} provides a
natural framework for constructing generalizations of usual quantum theory
that incorporates medium decoherence. Generalizations suitable for quantum
gravity have been considered by a number of authors ({\it e.g.}
\cite{Har95c,Savup}). 
But generalizations based upon extended probabilities would provide even
greater scope. Indeed a general quantum mechanical theory could be
specified just by giving a real valued function for candidate probabilities on
the sets of fine-grained histories. 

``Cheshire Puss [said Alice]\dots
would you tell me please, which way should I go from here? That
depends a good deal on where you want to get to, said the cat''. 

\begin{acknowledgments}
Conversations with Murray Gell-Mann over a long period of time were
helpful. Thanks are due to Todd Brun, Shelly Goldstein,
Jonathan Halliwell, Mark Srednicki, and Gary Horowitz for
useful discussions. The work was supported in part by the National Science 
Foundation under grant PHY00-70895.
\end{acknowledgments}

\appendix*
\renewcommand{\theequation}{\Alph{section}.\arabic{equation}}

\section{Eigenvalues of a Product of Projections}

The Hermitian part of the product of two non-commuting projections has at
least one negative eigenvalue. This appendix gives a proof of this
undoubtedly well-known fact that the author learned from G.T.~Horowitz. 

Let $P_a$ and $P_b$ be the two non-commuting projections and let $G\equiv P_aP_b
+ P_b P_a$ denote their Hermitian product. Let $\lambda_i$ and $|i\rangle$
denote the eigenvalues and eigenvectors of $G$. The expectation value in a
state $|\Psi\rangle$ is
\begin{equation}
\langle\Psi|G|\Psi\rangle = \sum\nolimits_i \lambda_i|\langle
i|\Psi\rangle|^2.
\label{aone}
\end{equation}
If there is one vector $|\Psi\rangle$ for which $\langle\Psi|G|\Psi\rangle$
is negative then $G$ must have one negative eigenvalue since otherwise
\eqref{aone} is positive. We now construct such a vector.

Any vector $|\Psi\rangle$ can be divided into orthogonal vectors
$|\Psi_1\rangle$ and $|\Psi_0\rangle$ that lie in the subspace $P_a$ and
the subspace orthogonal to it, {\it viz.}
\begin{equation}
|\Psi\rangle= |\Psi_1\rangle + |\Psi_0\rangle,\qquad P_a|\Psi_0\rangle=0,
\qquad P_a|\Psi_1\rangle=|\Psi_1\rangle.
\label{atwo}
\end{equation}
We now show how to pick $|\Psi_0\rangle$ and $|\Psi_1\rangle$ to make
$\langle \Psi|G|\Psi\rangle$ negative.

With the decomposition \eqref{atwo}
\begin{equation}
\langle\Psi|G|\Psi\rangle =
2\left[\left\langle\Psi_1\left|P_b\right|\Psi_1\right\rangle + {\rm Re}
\ \left\langle\Psi_1\left|P_b\right|\Psi_0\right\rangle\right].
\label{athree}
\end{equation}
For given $|\Psi_1\rangle$, $P_b|\Psi_1\rangle$ cannot be orthogonal to
every vector in $P_a$, or $P_b$ would commute with $P_a$ contrary to
assumption. Therefore pick $|\Psi_0\rangle$ so that ${\rm Re}
\ \langle\Psi_1|P_b|\Psi_0\rangle$ is non-vanishing.  If ${\rm Re}
\ \langle\Psi_1|P_b|\Psi_0\rangle >0$ replace $|\Psi_0\rangle$ by
$-|\Psi_0\rangle$ so that ${\rm Re}\ \langle\Psi_1|P_b|\Psi_0\rangle <0$. Now
make $|\Psi_0\rangle$ large enough that the negative second term in
\eqref{athree} is larger than the positive first term. The result is a
$|\Psi\rangle$ such that $\langle\Psi|G|\Psi\rangle \leq 0$.

The case of two one-dimensional projections
\begin{equation}
P_a= |a\rangle\, \langle a| \qquad P_b= |b\rangle\, \langle b|
\label{afour}
\end{equation} 
but with $\langle a|b\rangle \not= 0$ gives a concrete illustration of the
above result. The the eigenvectors of $G$ lie in the two-dimensional space
spanned by $|a\rangle$ and $|b\rangle$ and are easily calculated.
The two eigenvalues are  
\begin{equation}
\lambda_\pm = c^2 \pm c\ , \ {\rm where}\ \ c\equiv|\langle a|b\rangle|.
\label{afive}
\end{equation}
Since $c\leq 1$, $\lambda_-$ is evidently negative unless $c=0$ or $c=1$.
In either of these cases the projectors commute.


\begin{thebibliography}{99}

\bibitem{Gri02} R.B.~Griffiths, {\sl Consistent Quantum Theory}, Cambridge
University Press, Cambridge (2002).

\bibitem{Omn94} R.~Omn\`es, {\sl Interpretation of Quantum
Mechanics}, Princeton University Press, Princeton (1994).

\bibitem{Gel94} M.~Gell-Mann, {\sl The Quark and the Jaguar}, 
W.~Freeman, San Francisco (1994)
 
\bibitem{GH90b} M.~Gell-Mann and J.B.~Hartle, {\it Alternative Decohering
Histories in Quantum Mechanics}, in the {\sl Proceedings of
the 25th International Conference on High Energy Physics, Singapore,
August, 2-8, 1990}, ed.~by K.K.~Phua and Y.~Yamaguchi, South East Asia
Theoretical Physics Association and Physical Society of Japan, distributed
by World Scientific, Singapore (1990).

\bibitem{GP95} S.~Goldstein and D.~Page, {\it Linearly Positive Histories
-- Probabilities for a Robust Family of Sequences of Quantum Events},
{\sl Phys.~Rev.~Lett.} {\bf 74}, 3715 (1995); gr-qc/9403055.

\bibitem{Har93a} For an introductory review in the present notation, see
J.B.~Hartle, {\it The Quantum Mechanics of Closed
Systems}, in {\sl Directions in General Relativity,
Volume 1: A Symposium and Collection of Essays in honor of Professor
Charles W. Misner's 60th Birthday}, ed. by B.-L.~Hu,
M.P.~Ryan, and C.V.~Vishveshwara, Cambridge University Press, Cambridge
(1993); gr-qc/9210006.

\bibitem{Dio94} L.~Diosi, {\it Critique of Weakly Decohering and
Linearly Positive Histories}, unpublished; gr-qc/9409017. 

\bibitem{HLM95} J.B.~Hartle, R.~Laflamme, and D.~Marolf,
{\it Conservation Laws in the Quantum Mechanics of Closed Systems},
{\sl Phys.~Rev.~D} {\bf 51},  7007 (1995); gr-qc/9410006.

\bibitem{Har95c} J.B.~Hartle, {\it Spacetime Quantum Mechanics and the
Quantum Mechanics of Spacetime} in {\sl
Gravitation and Quantizations}, Proceedings of the 1992 Les Houches
Summer School, edited by B. Julia and J. Zinn-Justin, Les Houches Summer
School Proceedings Vol.~LVII, North Holland, Amsterdam (1995);
gr-qc/9304006. A pr\'ecis of these lectures is given in J.B.~Hartle, {\it
Quantum Mechanics at the Planck Scale},
talk given at the {\sl Workshop on Physics at the Planck Scale}, Puri,
India, December 1994; gr-qc/9508023.

\bibitem{probsum} D.~Finkelstein, {\sl Trans.~N.Y.~Acad.~Sci.} {\bf 25},
621 (1963); J.B.~Hartle, {\it Quantum Mechanics of Individual Systems},
{\sl Am.~J.~Phys.} {\bf 36}, 704 (1968); R.N.~Graham, in {\sl The Many-Worlds 
Interpretation of Quantum
Mechanics}, ed.~by B.~DeWitt and R.N.~Graham, Princeton University Press,
Princeton (1973); E.~Farhi, J.~Goldstone, and S.~Gutmann, {\sl Annals
of Phys.} (N.Y.) {\bf 192}, 368 (1989); Y.~Ohkuwa, {\it Decoherence Functional 
and Probability
Interpretation}, {\sl Phys.~Rev.~D} {\bf 48}, 1781 (1993); S.~Coleman and
A.~Lesniewski, unpublished; S.~Gutmann, {\it Using Classical Probability to
Guarantee Probabilities of Infinite Quantum Sequences}, {\sl Phys.~Rev.~A}
{\bf 52}, 3560 (1995); quant-ph/9506016

\bibitem{GH93b} M.~Gell-Mann and J.B.~Hartle, {\it Time Symmetry and
Asymmetry in Quantum Mechanics and Quantum Cosmology},
in {\sl Proceedings of the NATO Workshop on the Physical
Origins of Time Asymmetry, Mazag\'on, Spain, September 30-October4, 1991}
ed.~by J.~Halliwell, J.~P\'erez-Mercader, and W.~Zurek, Cambridge
University Press, Cambridge (1993); gr-qc/9309012.


\bibitem{GH90a} M.~Gell-Mann and J.B.~Hartle, {\it Quantum Mechanics in the
Light of Quantum Cosmology}, in {\sl Complexity, Entropy,
and the Physics of Information, SFI Studies in the Sciences of
Complexity}, Vol.~VIII, ed.~by W.~Zurek,  Addison Wesley, Reading, MA
(1990).

\bibitem{Har94b} J.B.~Hartle, {\it Quasiclassical Domains In A Quantum
Universe},
 in {\sl Proceedings of the Cornelius  Lanczos International
 Centenary Conference}, North Carolina State University,
December 1992, ed.~by J.D.~Brown, M.T.~Chu, D.C.~Ellison, R.J.~Plemmons,
SIAM, Philadelphia, (1994); gr-qc/9404017.

\bibitem{GH93a} M.~Gell-Mann and J.B.~Hartle, {\it Classical Equations for
Quantum Systems}, {\sl Phys.~Rev.~D} {\bf 47}, 3345 (1993); gr-qc/9210010.

\bibitem{GHup} M.~Gell-Mann and J.B.~Hartle, unpublished.

\bibitem{Lan61} R.~Landauer, {\sl IBM J.~Res.~Develop.} {\bf 3}, 183
(1961).

\bibitem{FH65} R.P.~Feynman and A.~Hibbs, {\sl Quantum Mechanics and Path
Integrals}, McGraw-Hill, New York (1965).

\bibitem{DH92} H.F.~Dowker and J.~Halliwell, {\sl Phys.~Rev.~D} {\bf 46},
1580 (1992).

\bibitem{AV91} Y.~Aharonov and L.~Vaidman, {\it Complete Description of a
Quantum System at a Given Time}, {\sl J.~Phys.~A} {\bf 24} 2315,
(1991).

\bibitem{Ken97} A.~Kent, {\it Consistent Sets Yield Contrary Influences in
Quantum Theory}, {\sl Phys.~Rev.~Lett.} {\bf 78}, 2874 (1997);
gr-qc/9604012.

\bibitem{GrH97} R.B.~Griffiths and J.B.~Hartle, Comment on {\it Consistent
Sets Yield Contrary Inferences in  Quantum Theory}, {\sl Phys.~Rev.~Lett.}
{\bf 81}, 1981 (1998); gr-qc/9710025.

\bibitem{Har91b} J.B.~Hartle, {\it Spacetime Coarse Grainings in
Non-Relativistic Quantum Mechanics}, {\sl Phys.~Rev.~D} {\bf 44}, 3173
(1991).

\bibitem{YT91} N.~Yamada and S.~Takagi, {\sl Prog.~Theor.~Phys.} {\bf 85},
985 (1991); {\it ibid.}, {\sl Prog.~Theor.~Phys.} {\bf 86},
599 (1991); {\it ibid.}, {\sl Prog.~Theor.~Phys.} {\bf 87},
77 (1992).

\bibitem{Cav86}  C.~Caves, {\it Quantum Mechanics of Measurements
Distributed in Time: A Path Integral Formulation},
{\sl Phys.~Rev.D} {\bf 33}, 1643 (1986).

\bibitem{MH96} R.J.~Micanek and J.B.~Hartle,
{\it Nearly Instantaneous Alternatives in Quantum
Mechanics}, {\sl Phys.~Rev.~A} {\bf 54},
3795-3800 (1996); quant-ph/9602023.

\bibitem{Har91a} J.B.~Hartle, {\it The Quantum Mechanics
of Cosmology}, in {\sl Quantum Cosmology and Baby Universes: 
Proceedings of the 1989 Jerusalem Winter School for Theoretical Physics},
ed.~by ~S.~Coleman, J.B.~Hartle, T.~Piran, and S.~Weinberg,
World Scientific, Singapore (1991) pp.~65-157, Section II.10.

\bibitem{Har98b} J.B.~Hartle, {\it Quantum Pasts and the
Utility of History}, {\sl Physica Scripta} {\bf T76}, 67--77 (1998);
gr-qc/9712001.

\bibitem{Fey87} R.P.~Feynman, {\it Negative Probability} in
{\sl Quantum Implications: Essays in Honor
of David Bohm}, ed.~by B.J.~Hiley and F.D.~Peat, Routledge and Kegan Paul,
London (1987).

\bibitem{fnnine} References
\cite{Har94b,GH93a,Omn94,GH95,Ana97,Zur03,BH99} illustrate the diversity of
approaches to defining
classicality. They are generally not the earliest or the latest.

\bibitem{GH95} M.~Gell-Mann and J.B.~Hartle, {\it Strong Decoherence},
in the {\sl Proceedings of the 4th Drexel Symposium on
Quantum Non-Integrability --- The Quantum-Classical Correspondence},
Drexel University, September 8-11, 1994, ed.~by D.-H.~Feng and B.-L.~Hu,
International Press, Boston/Hong-Kong; gr-qc/9509054. 

\bibitem{Ana97} C.~Anastopoulos, {\it Information Measures and
Classicality in Quantum Mechanics}, {\sl Phys.~Rev.~D} {\bf 59},
045001 (1999); quant-ph/9805005.

\bibitem{Zur03} W.~Zurek, {\it Decoherence, Einselection and the
Quantum Origin of the Classical}, {\sl Rev. Mod. Phys.} {\bf 75}, 75
(2003).

\bibitem{BH99} T.~Brun and J.B.~Hartle, {\it Classical Dynamics of the
Quantum Harmonic Chain}, {\sl Phys.~Rev.~D} {\bf 60}, 123503 (1999);
quant-ph/9905079.

\bibitem{JZ85} E.~Joos and H.D.~Zeh, {\sl Zeit.~Phys.} {\bf B59}, 223
(1985).

\bibitem{CL83} A.~Caldeira and A.~Leggett, {\it Path Integral Approach to
Quantum Brownian Motion}, {\sl Physica} {\bf 121A}, 587 (1983).

\bibitem{Zur84} W.~Zurek, in {\sl Non-Equilibrium Quantum
Statistical Physics}, ed.~by G.~Moore and M.~Scully,  Plenum Press,
New York (1984); quant-ph/0302044.  

\bibitem{Hal99} J.~Halliwell, {\it Somewhere in the Universe: Where is the
Information Stored When Histories Decohere}, {\sl Phys.~Rev.~D} {\bf 60},
105031 (1999); quant-ph/9902008.

\bibitem{Hal03} J.~Halliwell,
{\it Decoherence of Histories and Hydrodynamic Equations
for Linear Oscillator Chain} (2003); quant-ph/0305084.

\bibitem{DK96} H.F.~Dowker and A.~Kent, {\it On the Consistent Histories
Approach to Quantum Mechanics}, {\sl J.~Stat.~Phys.} {\bf 82}, 1574 (1996);
gr-qc/9412067.

\bibitem{Ish94} C.J.~Isham, {\it Quantum Logic and the Histories Approach
to Quantum Theory}, {\sl J.~Math.~Phys.} {\bf 35}, 2157
(1994); quant-ph/9308006.

\bibitem{IL94} C.J.~Isham and N.~Linden, {\it Quantum Temporal Logic
and Decoherence Functionals in the Histories Approach to Generalized
Quantum Theory}, {\sl J.~Math.~Phys.}, {\bf 35}, 5452
(1994); gr-qc/9405029.

\bibitem{Savup} K.~Savvidou, {\it General Relativity Histories Theory I:
The Spacetime Character of the Canonical Description}; gr-qc/030634. {\it
General Relativity Histories Theory II: Invariance Groups}; gr-qc/0306036.




\end{thebibliography}
\end{document}